\documentclass[prd,aps,twocolumn,a4paper,showkeys,nofootinbib]{revtex4-1}

\usepackage{graphicx,psfrag}
\usepackage{mathrsfs}
\usepackage{amsmath,amsfonts,amssymb}
\usepackage{multirow}
\usepackage{comment}
\usepackage{ulem} \usepackage{multirow}
\usepackage{hyperref}

\newcommand{\be}{\begin{equation}}
\newcommand{\ee}{\end{equation}}
\newcommand{\bea}{\begin{eqnarray}}
\newcommand{\eea}{\end{eqnarray}}
\newcommand{\bel}{\begin{align}}
\newcommand{\eel}{\end{align}}

\def\half{\frac{1}{2}}
\def\e{{\rm e}}
\def\i{{\rm i}}
\def\d{{\rm d}}

\def\Msun{{\rm M_{\odot}}}

\def\F{{{\mathcal{F}}}}
\def\Fbar{{\bar{\mathcal{F}}}}
\def\fitparams{{\params_{\rm fit}}}
\def\freeparams{{\params_{\rm free}}}

\def\twshift{{\tau_0}}
\def\kt{\kappa^{\rm T}_2}

\def\dkl{D_{\text{KL}}}
\def\params{\boldsymbol{\theta}}
\def\recalib{\boldsymbol{\delta}}
\def\recalibpm{{\recalib_{\rm fit}}}
\def\tsfzero{{T_0}}
\def\GMc2{{\rm G M_{\odot} c^{-2}}}

\def\wavelet{{W}}
\def\wavefm{{\tilde W}}
\def\waveam{{\breve W}}
\def\wavegen{{\widetilde W}}
\def\fm{{F}}
\def\fmzero{{F_0}}
\def\core{{\scshape CoRe}}
\def\sacra{{\scshape SACRA}}
\def\model{{\tt NRPMw}}
\def\oldmodel{{\tt NRPM}}

\def\rea{\Re{(\alpha)} }
\def\ima{\Im{(\alpha)} }
\def\reb{\Re{(\beta)} }
\def\imb{\Im{(\beta)} }
\def\tLam{{\tilde\Lambda}}

\def\spin{{\boldsymbol \chi}}
\def\chieff{{\chi_{\rm eff}}}
\def\adrift{\alpha_{\rm peak}}

\newcommand{\realabel}[1]{{ \Re{(\alpha_{#1})}}}
\newcommand{\imalabel}[1]{{ \Im{(\alpha_{#1})}}}
\newcommand{\reblabel}[1]{{ \Re{(\beta_{#1})}}}
\newcommand{\imblabel}[1]{{ \Im{(\beta_{#1})}}}
\newcommand{\xtt}[1]{{ \times 10^{{#1}}}}

\usepackage{pifont} 

\usepackage{color}
\definecolor{cyan}{rgb}{0,0.9,0.9}
\definecolor{orange}{rgb}{0.9,0.5,0}
\definecolor{magenta}{rgb}{1,0,1}
\definecolor{purple}{rgb}{0.8,0.4,0.8}
\definecolor{gray}{rgb}{0.8242,0.8242,0.8242}
\definecolor{cadmiumgreen}{rgb}{0.0, 0.42, 0.24}

\begin{document}

\title{Kilohertz Gravitational Waves From Binary Neutron Star Mergers:\\
  Numerical-relativity Informed Postmerger Model} 

\author{Matteo \surname{Breschi}$^{1}$}
\author{Sebastiano \surname{Bernuzzi}$^{1}$}
\author{Kabir \surname{Chakravarti}$^{1}$}
\author{Alessandro \surname{Camilletti}$^{2,3}$}
\author{Aviral \surname{Prakash}$^{5,6}$}
\author{Albino \surname{Perego}$^{2,3}$}

\affiliation{${}^{1}$Theoretisch-Physikalisches Institut, Friedrich-Schiller-Universit{\"a}t Jena, 07743, Jena, Germany}
\affiliation{${}^{2}$Dipartimento di Fisica, Universit\`{a} di Trento, Via Sommarive 14, 38123 Trento, Italy}
\affiliation{${}^{3}$INFN-TIFPA, Trento Institute for Fundamental Physics and Applications, via Sommarive 14, I-38123 Trento, Italy}
\affiliation{${}^{5}$Institute for Gravitation \& the Cosmos, The Pennsylvania State University, University Park, PA 16802, USA}
\affiliation{${}^{6}$Department of Physics, The Pennsylvania State University, University Park, PA 16802, USA}
\date{\today}

\begin{abstract}
  We present {\model}, an analytical model of gravitational-waves from 
  neutron star merger remnants informed using 618 numerical
  relativity (NR) simulations. 
  {\model} is designed in the frequency domain using a combination of
  complex Gaussian wavelets. The wavelet's parameters are calibrated to 
  equations of state (EOS) insensitive relations from NR data. 
  The NR simulations are computed with 21 EOS (7 of which are finite-temperature 
  microphysical models, and 3 of which contain quark phase
  transitions or hyperonic degrees of freedom) and 
  span total binary masses $M\in[2.4,3.4]~\Msun$,
  mass ratios up to $q=2$, and (nonprecessing) dimensionless spins magnitudes up to ${0.2}$.
  The theoretical uncertainties of the EOS-insensitive relations are incorporated in {\model} using recalibration parameters that enhance the flexibility and accuracy of the model.  
  {\model} is NR-faithful with fitting factors ${\gtrsim}0.9$
  computed on an independent validation set of $102$ simulations. 
\end{abstract}


\maketitle

\section{Introduction}
\label{sec:intro}

This work is the first of a series of papers that present a
faithful and complete (inspiral-merger-postmerger) model for
gravitational-wave (GW) signals from binary neutron star (BNS)
mergers, and its application to GW analyses with the third-generation 
Einstein Telescope (ET) detector~\cite{Hild:2008ng,Hild:2010id,
Hild:2011np,Punturo:2010zz,Maggiore:2019uih}. 
Our model builds on a state-of-art effective-one-body (EOB) 
approach for the inspiral-merger regime
\cite{Bernuzzi:2014owa,Nagar:2018zoe,Akcay:2020qrj,Gamba:2020ljo} 
and on its numerical relativity (NR) completion for  
the remnant's emission \cite{Bernuzzi:2015rla,Breschi:2019srl}. 
Prospects applications to ET GW observations 
include: the precision measurement of the neutron star (NS) tidal
polarizability parameters \cite{Damour:2009vw,Damour:2012yf}, 
the determination of the remnant's black hole (BH) collapse \cite{Agathos:2019sah,Kashyap:2021wzs}, 
constraints on the extreme density equation of state (EOS) \cite{Prakash:2021wpz,Breschi:2021xrx},
and multi-messenger observations \cite{Breschi:2021tbm}.
These case studies will be further discussed in companion papers in the context of a Bayesian
analysis framework \cite{BreschiInPrep}. 
Here, we start presenting {\model}, a new analytical model for the
postmerger (PM) emission from merger's remnant, that improves over
our previous {\oldmodel}~\cite{Breschi:2019srl}. 

The PM GW emission from a merger's remnant is predicted to
have a peak luminosity at frequencies of few kilohertz, e.g.~\cite{Shibata:2006nm,Hotokezaka:2013iia,Takami:2014tva,Bauswein:2015yca,Zappa:2017xba,Breschi:2019srl}.
This high-frequency GW transient can be robustly computed by means of  
NR simulations and it is key to directly
probe the nature of the remnant in a (possibly multi-messenger) BNS merger observation. A GW observation from a merger remnant is also a promising probe for the nuclear
EOS at extreme densities, e.g.~\cite{Radice:2016rys,Bauswein:2018bma,Breschi:2021xrx,Most:2021ktk,Prakash:2021wpz}.
Kilohertz PM transients are unlikely to be captured by current ground-base detectors~\cite{Aasi:2013wya}, and no PM signal was
detected for GW170817 \cite{TheLIGOScientific:2017qsa,Abbott:2017dke,Abbott:2018hgk,Abbott:2018wiz}.
However, they are a main target for third-generation observatories~\cite{Hild:2010id,Punturo:2010zza,Torres-Rivas:2018svp,Maggiore:2019uih,Martynov:2019gvu}
and for finely-tuned instruments~\cite{Ackley:2020atn}.
In view of these considerations, it is essential to develop accurate PM models for Bayesian GW analyses.

Models of PM GWs were presented in Refs.~\cite{Hotokezaka:2013iia,Takami:2014tva,Clark:2015zxa,Chatziioannou:2017ixj,Easter:2018pqy,Bose:2017jvk,Breschi:2019srl,Tsang:2019esi,Easter:2020ifj,Soultanis:2021oia,Wijngaarden:2022sah,Whittaker:2022pkd}. These templates are phenomenological models that capture the main PM spectral features but do not attempt to model the underlining remnant's dynamics. The complex spectral frequencies are either inferred from the observations or (in part) fixed by EOS-insensitive (quasiuniversal) relations that connect the main spectral features to the binary parameters. Depending on whether the quasiuniversal relations are employed or not during the GW data inference (and for which quantities), the templates might be used in fully-informed, partially informed or agnostic approach. Importantly, {\it all} approaches require the quasiuniversal relations to extract astrophysical constraints, either a priori or a posteriori.

Most of the PM templates are built from a simple ansatz made of few damped sinusoids in the time domain, eventually represented in the frequency domain.
Notable exceptions to a sinusoids basis are the models proposed in Refs.~\cite{Clark:2015zxa,Easter:2018pqy} where reduced basis were constructed directly from NR data. \citet{Clark:2015zxa} used a principal component analysis and ${\sim}50$ non-spinning simulations (12 of which unequal masses) to demonstrate faithfulnesses ${\gtrsim}0.9$ on a subsample of the data. \citet{Easter:2018pqy} used a hierarchical model trained on 35 non-spinning, equal-mass NR simulations to demonstrate fitting factors up to $0.98$ on the training set. However, similar fitting factors can be achieved with significantly less modeling efforts in agnostic approaches based on wavelets or sinusoids basis \cite{Chatziioannou:2017ixj,Tsang:2019esi,
	Easter:2020ifj,Wijngaarden:2022sah}. 
Moreover, the finite precision of NR simulations introduces uncertainties that impact the faithfulnesses at ${\sim}0.9$ level \cite{Easter:2018pqy,Breschi:2019srl}. 
Hence, simpler analytical templates appear favored over more complex statistical models. 
The agnostic approach utilized in Ref.~\cite{Chatziioannou:2017ixj,Wijngaarden:2022sah}
delivers, on average, larger fitting factors to numerical data when compared to fully or partially informed approaches, e.g.~\cite{Breschi:2019srl,Tsang:2019esi,Easter:2020ifj,Soultanis:2021oia}. This suggests that agnostic approaches are able to detect PM signals at lower signal-to-noise ratio (SNR) because informed models are not sufficiently accurate. The two approaches, however, appear comparable at SNR relevant for astrophysical parameter estimation, and they deliver comparable constraints on the EOS. We stress that faithfulnesses calculations are often presented on validation datasets of different sizes and a detailed comparison is difficult. For example, \citet{Easter:2020ifj} found faithfulnesses between 0.91-0.97 on a sample of 9 simulations; \citet{Tsang:2019esi} found faithfulnesses between 0.4-0.95 on a sample of 60 simulations, and \citet{Breschi:2019srl} between 0.4-0.95 on a sample of about $150$ simulations.
A main motivation for (partially) informed approaches is the possibility to design inspiral-merger-postmerger templates by consistently extending inspiral-merger templates. In Ref.~\cite{Breschi:2019srl}, we developed the first model of this kind by completing the EOB framework of Refs.~\cite{Bernuzzi:2014owa,Nagar:2018zoe} with the {\oldmodel} PM model.

The new {\model} is a PM frequency-domain template that aims at striking a balance between fully-informed and agnostic approaches. It is constructed by superposing few Gaussian, frequency-modulated wavelets whose parameters are informed by new EOS-insensitive relations. The latter build on the largest public databases of NR simulations available to date. The theoretical uncertainties of the EOS-insensitive relations are incorporated in the model using recalibration parameters that are determined during the inference. Hence, {\model} performs best in a partially informed inference. The recalibration enhances the flexibility of the template and improves the fitting factors to a level similar to agnostic templates. Data analysis applications of {\model} are presented in a companion paper.

The rest of this paper is structured as follows.
In Sec.~\ref{sec:phen}, we discuss the PM waveforms' phenomenology
predicted by state-of-art NR simulations.
The modeling choices used in {\model} are 
presented in Sec.~\ref{sec:model}.
The quasiuniversal relations calibrated for {\model} are discussed in
Sec.~\ref{sec:calib}. 
In Sec.~\ref{sec:valid} we validate the model against NR data by
calculating its faithfulness on an independent validation set.
We summarize our findings and conclude in Sec.~\ref{sec:conclusion}.
Moreover, 
we include several Appendices in order to extend
 the discussions on the waveform modeling
and on the calibration of EOS-insensitive relations.

\paragraph*{Conventions --}
We use geometric units $c=G=1$ or explicitely state units.
Masses are expressed in solar masses $\Msun$. 
The GW polarizations $h_+$ and $h_\times$,
 plus and cross respectively,
 are decomposed in $(\ell,m)$ multipoles as
\be
\label{eq:hdecomp}
h_+ - \i h_\times
=D_L^{-1}\sum_{\ell=2}^\infty\sum_{m=-\ell}^{\ell} h_{\ell m}(t)\,{}_{-2}Y_{\ell m}(\iota,\varphi),
\ee
where $D_L$ is the luminosity distance, 
${}_{-2}Y_{\ell m}$ are the $s=-2$
spin-weighted spherical harmonics
and $\iota,\varphi$ are respectively the polar and azimuthal 
angles that define the orientation of the binary with respect to the 
observer. 
Each mode $h_{\ell m}(t)$ is decomposed 
in amplitude $A_{\ell m}(t)$ and phase $\phi_{\ell m}(t)$, as
\be
\label{eq:hlm}
h_{\ell m}(t) = A_{\ell m}(t)\,\e^{- \i \phi_{\ell m}(t)} \,,
\ee
with a related GW frequency,
\be
\label{eq:fgw}
\omega_{\ell m}(t) =2\pi f_{\ell m}(t) = \frac{\d}{\d t}{\phi_{\ell m}}(t) \, .
\ee
The moment of merger is defined as the time of
the peak of $A_{22}(t)$, and referred simply as merger when it cannot
be confused with the coalescence/merger process.
If the multipolar indeces $(\ell, m)$ are omitted from a multipolar
quantity, we implicitly refer to the dominant $(2,2)$ mode.
Note that the time $t$ refers to the retarded time in the case of NR
data.
We define the Fourier transform $h_{\ell m}(f)$ of each multipolar mode as 
\be
\label{eq:fourier}
h_{\ell m}(f) = \int_{-\infty}^{+\infty} h_{\ell m}(t)\, \e^{-2\pi\i f t  }\,\d t \,.
\ee
Analogously to the time-domain case, 
the frequency series $h_{\ell m}(f)$ is decomposed  
in amplitude $A_{\ell m}(f)$ and phase $\phi_{\ell m}(f)$.

The binary mass is $M= m_1 +m_2$, where $m_{1,2}$ are the masses of the two stars, the mass ratio $q = m_1 /m_2 \ge 1$,
and the symmetric mass ratio $\nu = m_1 m_2 / M^2$.
We define the parameter $X =  1- 4 \nu $.
The dimensionless spin vectors are denoted with $\spin_i$ for $i=1,2$
and the spin component aligned with the orbital angular momentum
$\textbf{L}$ are labeled as $\chi_i = \spin_{i}\cdot \textbf{L} / |\textbf{L}|$.
The effective spin parameter $\chieff$ is the mass-weighted aligned spin, i.e.
\be
\label{eq:chieff}
\chieff = \frac{m_1 \chi_{1}+m_2 \chi_{2}}{M}\,.
\ee
Moreover,
the quadrupolar tidal deformability parameters are defined 
as $\Lambda_{i}=({2}/{3})\,k_{2,i}\,C_i^{-5}$ for $i=1,2$,
where $k_{2,i}$ and $C_i$ are respectively the $\ell=2$ gravitoelectric Love
number and the compactness of the $i$-th NS.
The tidal coupling constant is~\cite{Damour:2009wj}
\be
\label{eq:k2t}
\kt =3\nu\,\left[\left(\frac{m_1}{M}\right)^3 \Lambda_1 + (1\leftrightarrow 2)\right]\,,
\ee
that, similarly to the reduced tidal deformability $\tLam$~\cite{Favata:2013rwa},
parametrizes the leading-order tidal contribution to the binary interaction potential.

\section{Waveform morphology}
\label{sec:phen}

 \begin{figure}[t]
	\centering 
	\includegraphics[width=0.49\textwidth]{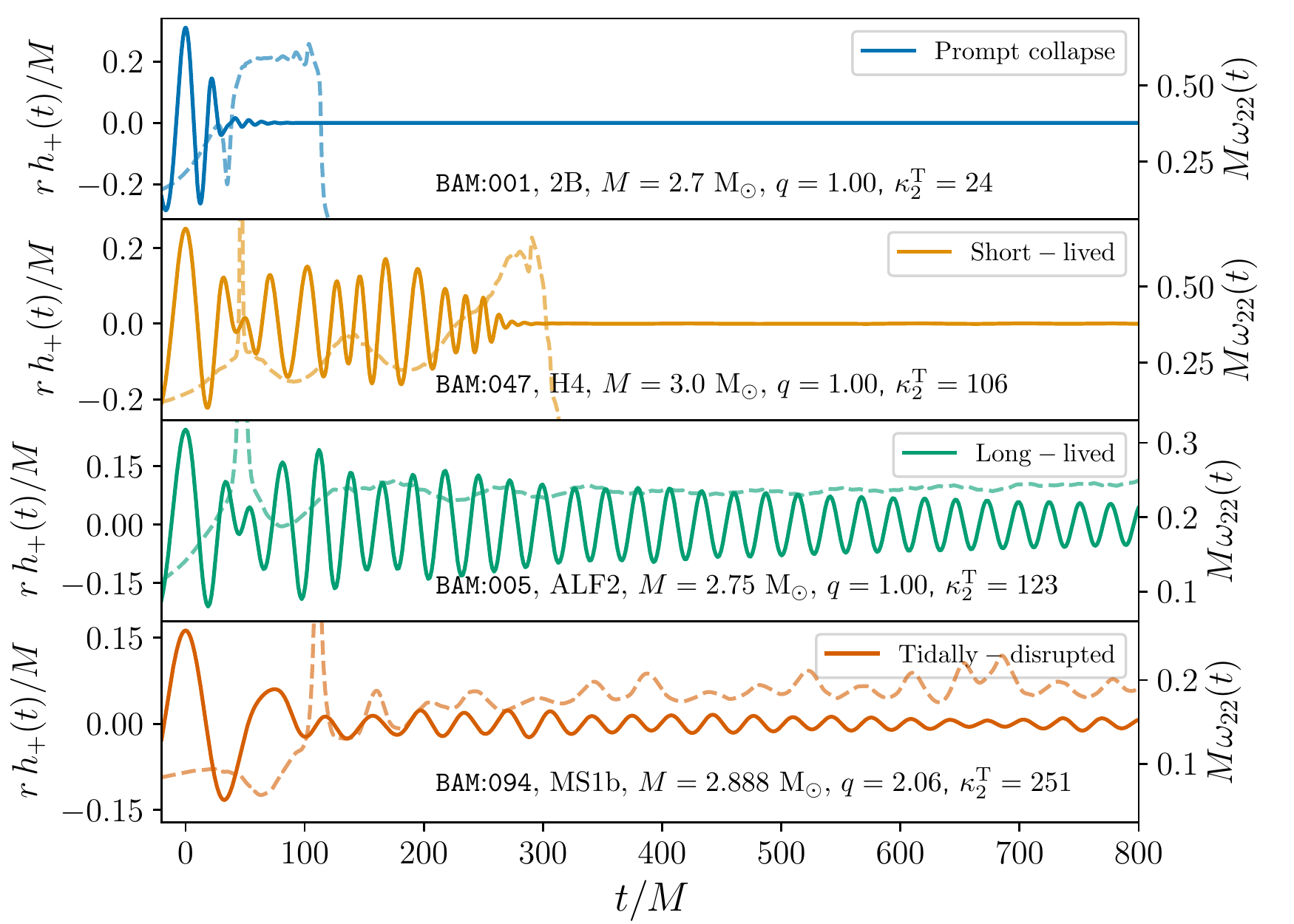}
	\caption{Representative examples of BNS PM waveforms.
          The plot shows the plus polarization $h_+(t)$ of the time-domain
          $\ell=m=2$ waveform (solid line)
          and the instantaneous GW frequency
          $\omega_{22}(t)$ (dashed line).
	  The NR simulations are from the {\core}
          database
          and computed in Refs.~\cite{Bernuzzi:2014owa,
          	Dietrich:2016hky}.
	}
	\label{fig:categories}
\end{figure}

The PM waveform morphology and its connection to the remnant's dynamics predicted by simulations was discussed in various papers, see e.g. Ref.~\citep{Shibata:2002jb,Stergioulas:2011gd,Bauswein:2011tp,Bauswein:2012ya,Hotokezaka:2013iia,Takami:2014zpa,Bauswein:2015yca,Bernuzzi:2015rla,Radice:2016gym,Lehner:2016lxy,Dietrich:2016hky,Dietrich:2016lyp,Zappa:2017xba,Bernuzzi:2020txg}.
We review here the main aspects that are relevant for the GW model proposed in the rest of the paper.
 Figure~\ref{fig:categories} shows the PM signal in exemplary cases; the time axis is shifted to the moment of merger.

 A merger remnant is a massive, hot and rotating NS whose mass is usually larger than the maximum mass sustained by a cold, isolated Tolmann-Oppenheimer-Volkoff (TOV) NS. It can either collapse to a BH or settle to a stable rotating NS on secular timescales. Gravitational collapse to BH takes place as the remnant reaches densities comparable to the TOV's maximum density \cite{Perego:2021mkd} since the remnant's core is very slowly rotating \cite{Kastaun:2016elu}.
 The remnant of a very massive BNS can promptly collapse after the moment of merger and crucially
 before the first bounce of the two cores~\cite{Radice:2020ddv,Bernuzzi:2020tgt}~\footnote{
This definition of prompt collapse implies negligible shocked dynamical ejecta because the bulk of this mass ejection comes precisely from the first core bounce \cite{Radice:2018pdn}. Since it directly connects to the main dynamical feature of the merger process (shock and bounce) and to related observables, it is preferable to other empirical definitions based on collapse time from the moment of merger.}. 

 In the case of a equal mass BNS, the prompt collapse is described by empirical relations relating the binary mass to the TOV maximum mass and compactness proposed in Ref.~\cite{Hotokezaka:2011dh,Bauswein:2013jpa} and refined in various works, e.g.~\cite{Agathos:2019sah,Kashyap:2021wzs}.
For very asymmetric BNS, the tidal disruption of the secondary drives the gravitational collapse \cite{Bernuzzi:2020txg} and it is mainly controlled by the incompressibility parameter of nuclear matter around the TOV maximum density \cite{Perego:2021mkd}. While a robust prompt collapse criterion is not known in these conditions \cite{Bernuzzi:2020txg,Bauswein:2020aag,Perego:2021mkd}, tidal disruption effects are of the order of current EOS effects in the equal-mass criterion, at least for mass-ratio $q\lesssim1.4$ \cite{Kashyap:2021wzs,Perego:2021mkd}. 
Prompt collapse mergers have the largest GW luminosities (at merger) \cite{Zappa:2017xba} but the PM signal is the rapidly damped ringdown of the BH and it is practically negligible for the sensitivities of current and next-generation
detectors. A prompt collapse signal is showed in the top panel of Figure~\ref{fig:categories}.

 The evolution of a NS remnant is driven by an intense emission of GWs lasting ${\sim}10{-}20$~milliseconds (GW-driven phase) \cite{Bernuzzi:2015opx,Zappa:2017xba}. During this phase, the remnant either collapses to BH ({\it short-lived} remnant) or settles to an approximately axisymmetric rotating NS ({\it long-lived} remnant)~\footnote{A commonly used terminology for short-lived remnant is hypermassive NS. This name is not appropriate for remnants since it refers to cold equilibrium. See \cite{Radice:2018xqa,Bernuzzi:2020tgt} and references therein.}. The GW-driven phase is associated to a luminous GW transient at frequencies ${\sim}2-4\,$kHz \cite{Shibata:2002jb,Stergioulas:2011gd,Bauswein:2011tp,Hotokezaka:2013iia,Takami:2014zpa,Bernuzzi:2015rla}. The spectrum of this transient is rather complex but has robust and well-studied features at a few characteristic frequencies. Most of the power is emitted in the $\ell=m=2$ GW mode at a nearly constant frequency $\omega_{22}(t)\approx 2\pi f_2$. Examples of $\ell=m=2$ waveforms for short- and long-lived remnants are shown in the three bottom panels of Figure~\ref{fig:categories}. The $f_2$ frequency is easily extracted from simulation data and it was shown to correlate with various binary quantities in a EOS-insensitive way, e.g.~\cite{Bauswein:2015yca,Takami:2014zpa,Bernuzzi:2015rla,Breschi:2019srl}. 
 
 We stress that the PM spectrum is not composed of a discrete set of frequencies: the presence of broad peaks with typical full width at half maximum (FWHM) of $300{-}600$~Hz is simply a consequence of the efficiency of the emission process. Indeed, inspection of the time-domain waveform's instantaneous frequency (see Fig.~\ref{fig:categories}) shows that $\omega_{22}(t)$ increases as the remnant becomes more compact and has a steep acceleration towards gravitational collapse~\footnote{Considering gauge-invariant energetics it is possible to associate to the remnant a dynamical frequency $\Omega$ such that $f_2 = \Omega/\pi$ and analogously for other modes.}, see e.g. Fig.~1 of \cite{Bernuzzi:2015rla}. Moreover, the instantaneous GW frequency has modulations with frequencies $f_0\sim O(1~{\rm kHz})$ that are stronger for remnants closer to collapse. These modulations are associated to the violent radial bounces of the remnant's core prior to collapse. 
 Other robust features of the spectrum are two secondary peaks at frequencies $f_{2\pm0}$,
   respectively at larger and smaller frequencies than $f_2$. These features are associated to hydrodynamical modes in the remnant, e.g.~\cite{Shibata:1999wm,Stergioulas:2011gd,Bernuzzi:2013rza} and have been was tentatively interpreted as nonlinear coupling between $f_2$ and $f_0$ \cite{Stergioulas:2011gd}, in analogy to perturbations of rotating NS \cite{Dimmelmeier:2005zk,Passamonti:2007tm,Baiotti:2008nf}.
 The remnant's signal from asymmetric binaries with mass ratio $q\gtrsim1.5$ carries the imprint of the tidal disruption during merger. An example is shown in the bottom panel of Figure~\ref{fig:categories}. The PM amplitude can be significantly smaller than in the equal-mass cases and the peaks at frequencies $f_{2\pm0}$ are typically suppressed.

The evolution of a NS remnant beyond the GW-driven phase is highly uncertain at present. It requires detailed simulations of viscous and nuclear processes on timescales beyond hundreds of milliseconds, for example to quantify precisely the mass accreting or outflowing the central object. NS remnants after the GW-driven phase have an excess of both gravitational mass and angular momentum when compared to equilibrium configuration with the corresponding baryon mass \cite{Radice:2018xqa,Nedora:2020pak}. Possible mechanisms to shed (part of) this energy are long-term GW instabilities \cite{Chandrasekhar:1970b,Friedman:1978b} including one-arm instabilities \cite{East:2015vix,Radice:2016gym}, that would lead to potentially detectable, long GW transients at ${\lesssim}1\,$kHz.
 
The PM model presented in the next sections describes the GW transient during the GW-driven phase and it builds on our previous work in Refs.~\cite{Bernuzzi:2015rla,Breschi:2019srl}. In particular, we devise new EOS-insensitive relations based on the tidal coupling constant $\kt$ and incorporate them in a partially informed model. We do not use empirical relations for modeling prompt collapse and instead design a model capable of inferring a generic collapse time from the observational data (but see e.g. Ref.~\cite{Agathos:2019sah} for an application of prompt collapse quasiuniversal realtions in data analysis context). Similarly, we account for the theoretical uncertainties of the EOS-insensitive using recalibration parameters inferred from the data.

\section{{\model} design}
\label{sec:model}

In order to develop an analytical 
NR-informed PM model for BNS mergers 
in the frequency-domain,
we first introduce a truncated 
complex Gaussian wavelet $\wavelet(t)$,
\be
\label{eq:tdwavelet}
\wavelet(t;\alpha,\beta,\gamma,\tau) = 
\begin{cases}
	\begin{split}
 \e^{\alpha t^2+\beta t +\gamma} \quad& {\rm if }\,\,t\in [0,\tau]\\
0\quad& {\rm otherwise}
	\end{split}
\end{cases}
\ee
where $\alpha,\beta,\gamma\in \mathbb{C}$
are time-independent parameters and the real interval 
$[0,\tau]$ defines the 
non-vanishing support of $\wavelet$.
The coefficients $\{\alpha,\beta,\gamma\}$ can be interpreted as follows:
$\rea$ and $\reb$ determine respectively the concavity
and the initial slope of the time-domain wavelet amplitude;
$\ima$ and $\imb$ define respectively the slope and the 
initial value of the time-domain frequency evolution; 
$\gamma$ is an overall factor determining initial amplitude and phase.

The frequency-domain wavelet $\wavelet(f)$ 
can be analytically computed from Eq.~\eqref{eq:tdwavelet}
using Gaussian integrals, 
\be
\label{eq:fdwavelet}
\wavelet(f) = \frac{\e^{\gamma}}{2} 
\sqrt{\frac{\pi}{\alpha}}\, \e^{-z^2}\,
\left[{\rm erfi}{\left( z + \sqrt{\alpha} \tau \right)}- {\rm erfi}{\left( z \right)} \right]\,,
\ee
where $z(f)$ encodes the frequency dependency,
\be
\label{eq:z_of_f}
	z(f)  = \frac{\beta-2\pi\i f}{2\sqrt{\alpha}}\,,
\ee
and ${\rm erfi}{(z)}$ is the imaginary error function.
For $\alpha =0$, Eq.~\eqref{eq:fdwavelet} is not defined and it is
directly replaced by a Lorentzian function. 
Moreover, a direct implementation of Eq.~\eqref{eq:fdwavelet} can lead
to floating point overflow fin a certain portion of the parameter space.
In these cases, we employ the analytical approximations discussed in App.~\ref{app:numerror}.
Furthermore, we introduce a global time-shift $\twshift$
in order to allow the wavelet to move on the time axis. 
The time-shift $\twshift$ changes the wavelet support to
$[\twshift, \tau+\twshift]$ and it is easily implemented by a unitary factor, i.e. 
$\wavelet(f;\twshift)= \wavelet(f)\,\e^{-2\pi\i f\twshift }$.

The wavelet is the basic component of {\model}.
In the following
paragraphs we describe how different wavelets are combined based on the
universal features of the PM signal that are identified by
characteristic times (``nodes'', Sec.~\ref{sec:nodes}). 
Then, we discuss the modeling of sub-dominant frequencies as additional wavelet modulations in
Sec.~\ref{sec:modulations}.
The basic construction of the dominant $\ell=m=2$ mode is discussed
Sec.~\ref{sec:dominant} and the modeling of higher-order modes in Sec.~\ref{sec:hms}.

\subsection{Nodal points}
\label{sec:nodes}

The time-domain strain has universal characteristic features at
specific times, as pointed out in Ref.~\cite{Breschi:2019srl} (see
also Fig.~2).
We call these times {\it nodal points} and indicate them as $\{t_i\}$ for $i=0,1,2,3$.
Nodal points are identified as stationary points of the strain's
amplitude, that we indicate as $\{A_i\}$.
Differently from Ref.~\cite{Breschi:2019srl}, we assume $t_{i+1}-t_i$ to be constant, 
for $i=0,1,2$. Hence, 
the nodal points can be reduced to two independent parameters: 
the moment $t_0$ of the first amplitude's minimum after
merger, and a characteristic time-scale $\tsfzero$ that is computed as the 
difference $t_{3}-t_1$.
The time-scale $\tsfzero$ defines the subdominant frequency $f_0 \simeq
\tsfzero^{-1}$ that characterizes the modulations 
of the PM signal.
A further time-domain node
  is introduced for the time of the remnant collapse $t_{\rm coll}$.
Differently from \cite{Breschi:2019srl},
  here we do not introduce $t_4$. 

\subsection{Amplitude and frequency modulations}
\label{sec:modulations}

Ampitude and frequency modulations (AMs, FMs)
are prominent features of the PM spectrum, as discussed in Sec.~\ref{sec:phen}.
NR simulations show that the main GW modulations are given in the $m=0$
channel, and are associated to the quasi-radial  
density oscillations of the remnant~\cite{Bernuzzi:2008rq}.
We associate this mode to the fundamental frequency $f_0$
and, for the modeling of the $(2,2)$ mode,
we consider only the modulation couplings between
$f_2$ and $f_0$
\footnote{Our simulations indicate that the couplings between $(2,1)$ and $(3,3)$ modes can also be relevant for unequal-mass BNS.}.
Moreover, we neglect possible frequency evolution of the subdominant 
mode $f_0$, i.e. this frequency component is assumed to be constant in time. 
Modulation effects appear after the collision of the NS cores, for $t>t_{0}$,
when the remnant is strongly deformed and dynamically unstable.

AMs can be easily taken into account by employing a combination of
wavelets. Labeling the amplitude-modulated wavelet as $\waveam$, 
we can write
\be
\label{eq:am}
\begin{split}
	\waveam(t) &= \wavelet(t) \left[ 1 + \Delta_{\rm am}  \sin\left( \Omega_{\rm am}t +\phi_{\rm am} \right) \right]\\
			&= \wavelet(t) -\frac{\i \Delta_{\rm am}}{2}\sum_{k=\pm 1} k\, \wavelet(t)\,\e^{\i k(\Omega_{\rm am}t +\phi_{\rm am})} \,,
\end{split}
\ee
where $\Delta_{\rm am}$ defines the magnitude,
$\Omega_{\rm am}$ the modulation frequency
and $\phi_{\rm am}$ the initial phase of the AMs.
Eq.~\eqref{eq:am} can be transformed in the Fourier space
obtaining
\be
\label{eq:amfd}
	\waveam(f) = \wavelet(f) -\frac{\i \Delta_{\rm am}}{2}\sum_{k=\pm 1} k\, \wavelet^{(k)}(f) \,,
\ee
where
\be
\label{eq:amfd_wk}
 \wavelet^{(k)}(f) = \wavelet(f; \alpha, \beta +\i k \Omega_{\rm am}, \gamma +\i k \phi_{\rm am}, \tau)  \,.
\ee
Eq.~\eqref{eq:amfd} 
shows explicitly that an amplitude-modulated wavelet 
$\waveam$ can be easily written in terms of the Gaussian wavelets $\wavelet$
and it introduces two subdominant contributions in the Fourier domain
that are displaced with respect to the primary peak of $\pm \Omega_{\rm am}$.

FMs affect the phase evolution
of the time-domain wavelet.
We implement a FM wavelet $\wavefm$
defining the frequency evolution as
\be
\label{eq:fhfm_main}
\omega_\wavefm(t) 
= \omega_\wavelet(t) - \Delta_{\rm fm} \e^{-\Gamma_{\rm fm} t}\sin(\Omega_{\rm fm} t + \phi_{\rm fm})\,,
\ee
where 
$\omega_\wavefm$ is the instantaneous frequency of the frequency-modulated wavelet $\wavefm$,
$\omega_\wavelet$ is the instantaneous frequency of the Gaussian wavelet $\wavelet$,
and $\Delta_{\rm fm},~\Gamma_{\rm fm},~\Omega_{\rm fm},~\phi_{\rm fm}\in\mathbb{R}$
are the parameters that define the FM, i.e.
$\Delta_{\rm fm}$ is the initial frequency displacement,
$\Gamma_{\rm fm}$ the inverse damping time,
$\Omega_{\rm fm}$ the modulation frequency
and $\phi_{\rm fm}$ the initial phase.
Using Taylor expansion,
the frequency-modulated wavelet $\wavefm$ can be rewritten 
in terms of the frequency-domain Gaussian wavelet $\wavelet$.
A detailed discussion on the analytic form of $\wavefm(f)$ is provided in App.~\ref{app:fmapprox}.
Note that, differently from the AMs shown in
Eq.~\eqref{eq:am}, the FM contribution
presented in Eq.~\eqref{eq:fhfm_main} 
includes damped behavior,
i.e. $\Gamma_{\rm fm}\ne 0$ a priori.
This term is needed to properly characterize
the different time-scales of the 
PM frequency components $f_2$ and $f_0$.

Combining the definitions of $\waveam$, Eq.~\eqref{eq:amfd},
and $\wavefm$ (see Eq.~\eqref{eq:fmwave_fd}),
it is possible to write a general modulated 
Gaussian wavelet, labeled as $\wavegen$.
We consider AMs over the interval $[t_0,t_3]$
and FM for $t>t_0$.
We fix the modulation frequencies to $\Omega_{\rm am}=\Omega_{\rm fm}=2\pi f_0$.
Then, the AM magnitude $\Delta_{\rm am}$ 
and phase $\phi_{\rm am}$ 
are fixed by the values of the GW amplitudes at the nodal points nodal points, 
i.e. $\{t_i,A_i\}$ for $i=1,2,3$.
The FM inverse damping time $\Gamma_{\rm fm} $ is assumed to
be identically zero 
for $t < t_1$; then, 
it is fixed to a constant positive value
calibrated on NR data (see Sec.~\ref{sec:calib}).
Furthermore, NR simulations show
that AMs and FMs approximately
fluctuate in opposite directions~\cite{Breschi:2019srl};
i.e. amplitude maxima occur at frequency minima and viceversa.
The FM phase $\phi_{\rm fm}$ is fixed in order 
to satisfy this requirement.

\subsection{Wavelet combination}
\label{sec:dominant}

 \begin{figure}[t]
	\centering 
	\includegraphics[width=0.5\textwidth]{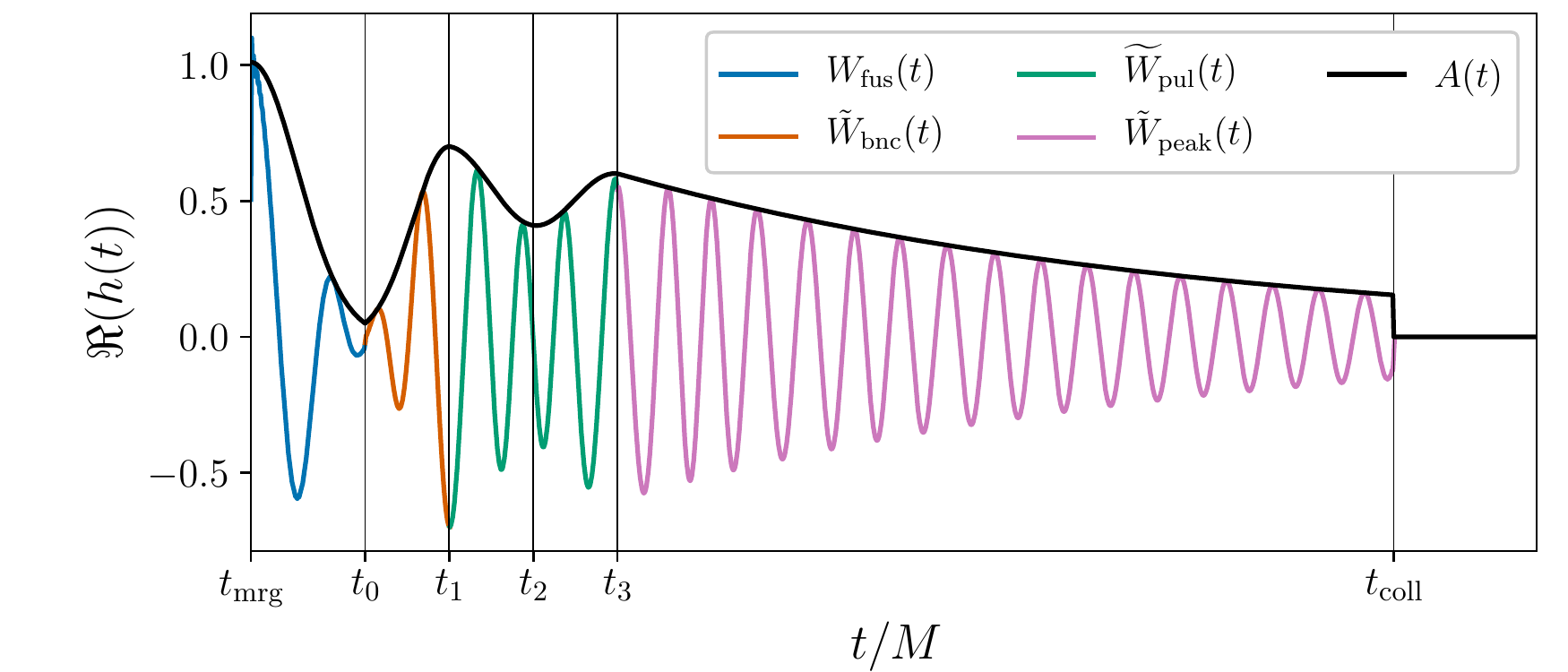}
	\includegraphics[width=0.5\textwidth]{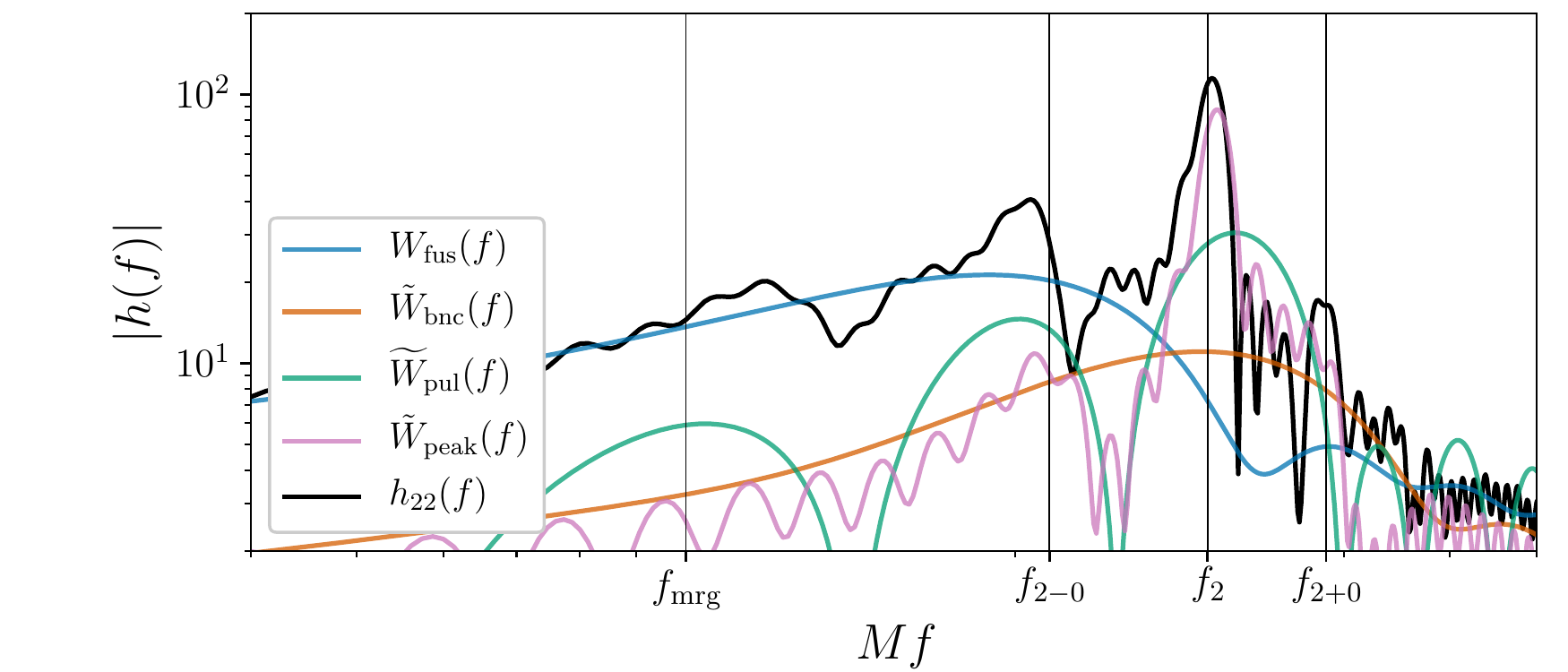}
	\caption{Exemplary case showing the morphology of {\model} model.
		Different wavelet components are reported with different colors:
		$\wavelet_{\rm fus}$ in blue, 
		$\wavefm_{\rm bnc}$ in orange,
		$\wavegen_{\rm pul}$ in green,
		and $\wavefm_{\rm peak}$ in purple.
		The top panel shows the time-domain
		components and the overall GW amplitude $A(t)$
		(black line)
		highlighting the characteristic times with vertical lines, i.e.
		the time of the merger $t_{\rm mrg}$, the nodal points $t_i$ for $i=0,1,2,3$
		and the time of collapse $t_{\rm coll}$.
		The bottom panel shows the Fourier spectra of each component, 
		the overall $h_{22}$ spectrum (black line) and 
		the characteristic PM frequencies (vertical lines),
		i.e. the merger frequency $f_{\rm mrg}$,
		the PM peak $f_{2}$ and 
		the subdominant couplings $f_{2\pm 0} = f_2\pm f_0$.
	}
	\label{fig:basic}
\end{figure}

The {\model} model is constructed by describing each part of the PM
signal between different nodal points with a
modulated wavelet component.
The overall strain $h_{22}$ is computed summing all the contributions.
The use of wavelets allow us to assign a clear interpretation of each parameter 
employed in the model.
The combination of different wavelets can capture rather complex signal morphologies.

In {\model}, the physical quantities (times, amplitudes and
frequencies) are estimated using quasiuniversal relations calibrated
on NR simulations (see Sec.~\ref{sec:calib}). 
This allows us to design a fully informed model that can connect the
signal's morphology to the intrinsic parameters of the BNS 
system (masses, spins and tidal parameters).
However, some wavelet parameters could be let unconstrained and directly inferred 
from observational data~\cite{Easter:2020ifj}
or they could be reconstructed with regression methods
directly from NR simulations~\cite{Easter:2018pqy}.

The time-domain $\ell=m=2$ mode is modeled employing a combination of
four different wavelet components,
\be
 \label{eq:h22}
 h(t)\approx \wavelet_{\rm fus}(t)+\wavefm_{\rm bnc}(t)+\wavegen_{\rm pul}(t)+\wavefm_{\rm peak}(t)\,,
 \ee
assuming continuity in amplitude and phase
(except for a phase-shift $\phi_{\rm PM}$, see later)
for the time-domain counterpart.
Detailed expressions are given in
App.~\ref{app:wavecomponent}.
The combination of wavelets includes
the following terms that are shown in color in Fig.~\ref{fig:basic}:
\begin{enumerate}
\item $\wavelet_{\rm fus}$ describes the signal after merger and up
  to $t_0$, corresponding to the fusion of the 
  NS cores.
  The wavelet has an initial frequency 
  $f_{\rm mrg}$
  and non-vanishing frequency drift that can be positive or negative
  depending on the properties of the binary;\\
\item $\wavefm_{\rm bnc}$ describes the signal corresponding to the
  bounce after the collision of the cores.
  The phase here has a discontinuity $\phi_{\rm PM}$ at $t_0$.
  Moreover, for $t>t_0$,
  all wavelets include FMs with the subdominant frequency $f_0$;\\
\item $\wavegen_{\rm pul}$ describes the emission up to $t_3$
  during which the remnant is typically highly dynamical. 
  Since the largest amount of the GW luminosity is emitted at
  times $\lesssim 5~{\rm ms}$~\cite{Zappa:2017xba},
  this component also includes AMs 
  with the subdominant frequency $f_0$;
\item $\wavefm_{\rm peak}$ describes the signal after the luminosity
  peak by a damped sinusoidal with initial frequency $f_2$, 
  a frequency evolution parametrized by the drift
  $\adrift$ (also referred as $\imalabel{\rm peak}$ in App.~\ref{app:wavecomponent}).
  This component characterizes the 
  dominant Fourier peak and it lasts 
  until the time of collapse $t_{\rm coll}$.
\end{enumerate}
  
Additionally, the GWs emitted by the collapse and BH ringdown can be
modeled as a fifth term in Eq.~\eqref{eq:h22}, $\wavelet_{\rm coll}$
(see App.~\ref{app:wavecomponent} for a detailed discussion).
Knowing the properties of the final BH,
this component could be modeled
with the quasi-normal modes 
of the remnant~\cite{Berti:2009kk,Berti:2014fga}.
For simplicity, however, we set here $\wavelet_{\rm coll}= 0$.
  
Figure~\ref{fig:basic} shows an example 
of the discussed contributions in time- and frequency-domain, with the
different terms appearing in Eq.~\ref{eq:h22} shown in different colors.
The overall spectrum shows the typical PM 
patterns:
a dominant quasi-Lorentzian peak,
a weaker peak at lower frequencies corresponding to the merger dynamics 
and sub-dominant peaks due to AMs and FMs.
The superposition of 
the wavelet components generates several local minima 
and maxima in the overall $h_{22}$ spectrum.
Moreover, the destructive interference of the wavelets 
originates a local minimum typically located between $f_{\rm mrg}$ and $f_2$.
This feature is also generally observed in BNS PM spectra extracted from NR simulations.
Moreover, the sharp cut at $t_{\rm coll}$ in time-domain waveform originates
the ringing effects 
observed in the $h_{22}$ spectrum
\footnote{This can be easily seen performing the convolution product of 
a sinusoidal wavelet with a Heaviside function.}.
The further inclusion of $W_{\rm coll}$ will mitigate
this effect, yielding to a smoother waveform representation.

Overall, the model is characterized by 
17 parameters,
that are the characteristic frequencies, amplitudes, times
and phases that define instantaneous GW amplitude and frequency
(see App.~\ref{app:wavecomponent}).
Most of these quantities can be related to the 
binary properties using NR information.

\subsection{Higher-order modes}
\label{sec:hms}

NR simulations show prominent
coupling effects in higher mode (HM) terms of BNS PM transients,
similarly to what we discussed for the dominant $(2,2)$ mode.
Also for this reason, the power of
HM contributions in BNS PM radiation is considerably
larger compared with the pre-merger dynamics~\cite{CalderonBustillo:2020kcg,Bernuzzi:2020txg}.
For typical BNS systems, these contributions
cover a relatively broad spectrum, roughly from ${\sim}500~$Hz
to 5--7~kHz.

In general, HM contributions can be modeled
as a combination of wavelets
with different frequencies imposing 
continuity in amplitude and phase.
For $m\ne 0$,
the characteristic peak frequencies of HMs can be 
approximated using the quadrupolar term
employing the multipolar scaling, i.e.
$f_{\ell m} \simeq ({m}/{2}) \, f_2$.
However, the hierarchy of frequency couplings is not fully resolved.
A detailed analysis of these subdominant features might require better resolved simulations to robustly identify the trend in the spectra.
We remand the inclusion of HM PM characteristic properties to a future study.

\section{NR calibration}
\label{sec:calib}

The {\model} model has 17 parameters, i.e.
\be
\label{eq:pmparams}
\begin{split}
\params_{\rm PM} = \{ &\phi_{\rm PM},\phi_{\rm fm},t_0, t_{\rm coll},\\
&A_{\rm mrg}, A_{0}, A_{1}, A_{2}, A_{3}, \\
&f_{\rm mrg}, f_{2}, f_{0}, \Delta_{\rm fm}, \Gamma_{\rm fm},\\
&\reblabel{\rm peak} ,\imalabel{\rm fus}, \adrift  \} \,,
\end{split}
\ee 
that can be mapped to the binary parameters,
\be
\label{eq:inspparams}
\params_{\rm bin}= \left\{ m_1, m_2, \Lambda_1, \Lambda_2,\chi_{1}, \chi_{2} \right\}\,,
\ee
using NR simulations.
We chose to map only a subset of $\params_{\rm PM}$ and let some other parameters to be determined by the inference or any other minimization procedure with given data.
In particular, we map the following 13 parameters
\be
\label{eq:fitparams}
\begin{split}
	\fitparams = \{ &A_{\rm mrg}, A_{0}, A_{1}, A_{2}, A_{3}, f_{\rm mrg}, f_{2}, f_{0},t_0,
	\\	&\reblabel{\rm peak} ,\imalabel{\rm fus},\Delta_{\rm fm},\Gamma_{\rm fm} \} \,,
\end{split}
\ee 
we fix $\phi_{\rm fm}$ by the AMs and the FMs 
as discussed in Sec.~\ref{sec:modulations},
and we leave three additional degrees of freedom,
\be
\label{eq:freeparams}
\freeparams= \left\{\phi_{\rm PM}, t_{\rm coll}, \adrift\right\}\,.
\ee
This choice is motivated by the fact that these three parameters cannot be robustly mapped using NR data. The PM phase $\phi_{\rm PM}$ shows a strong dependence on the simulation's grid resolutions
and on the physical models, e.g.~\cite{Bernuzzi:2011aq,Dietrich:2018upm,Bernuzzi:2016pie}.
The time of collapse $t_{\rm coll}$ is difficult to robustly determine
from simulations due its dependence on grid resolution \cite{Breschi:2019srl}; moreover, it strongly depends on the 
properties of the nuclear EOS and might be biased by the relatively small EOS set available~\cite{Radice:2016rys,Prakash:2021wpz,
Fujimoto:2022xhv}.
The frequency drift $\adrift$ is also connected to the collapse dynamics and, as such, it can be affected by various processes, especially in long-lived remnants. For example, we discuss in App.~\ref{app:viscosity} the dependency of $\adrift$ on the turbulent viscosity in a subset of simulations.

The calibration set of binaries includes the public available non-precessing NR simulations of the \core~\cite{Dietrich:2018phi, core_web} and the \sacra~\cite{Kawaguchi:2018gvj,Kiuchi:2017pte,Kiuchi:2019kzt} databases, plus additional data from simulations of Ref.~\cite{Bernuzzi:2020txg,Prakash:2021wpz,Camilletti:2022jms} with the BLh and BLQ EOS. The {\core} database includes data computed with two different NR codes, {\scshape BAM}~\cite{Brugmann:2008zz,Thierfelder:2011yi} and {\scshape THC}~\cite{Radice:2012cu} and simulate microphysics, neutrino transport (with various schemes) and turbulent viscosity. The final dataset is composed by 618 simulations and it includes 190 different binary configurations computed with three independent NR codes and 21 different EOSs. The finite temperature, composition-dependent EOSs are 
BHB$\Lambda\phi$~\cite{Banik:2014qja},
BLh~\cite{Bombaci:2018ksa,Logoteta:2020yxf},
BLQ~\cite{Bombaci:2018ksa,Logoteta:2020yxf,Prakash:2021wpz},
HS(DD2)~\citep[][DD2 hereafter]{Typel:2009sy,Hempel:2011mk},
LS220~\cite{Lattimer:1991nc},
SFHo~\cite{Steiner:2012xt},
SRO(SLy)~\citep[][SLy hereafter]{daSilvaSchneider:2017jpg};
the EOSs in piecewise polytropic forms are: 
ALF2~\cite{Alford:2004pf},
ENG~\cite{Engvik:1995gn},
MPA~\cite{Muther:1987xaa},
MS1~\cite{Mueller:1996pm},
MS1b~\cite{Mueller:1996pm},
SLy~\cite{Douchin:2001sv},
2B, 2H, 15H, 125H, B, H, H4, HB from Ref.~\cite{Lackey:2005tk,Read:2008iy}
and the $\Gamma{=}2$ ideal gas EOS.
We remark that ALF2 and BLQ include a phase transition to deconfined
quark matter, 
and BHB$\Lambda\phi$ takes into account the appearance of hyperons at high densities.
The intrinsic parameters of the data cover the ranges $M\in[2.4,3.4]~\Msun$, $q\in[1,2]$,
$\kt\in[22, 458]$ and $\chieff\in[-0.14,+0.22]$.
Among the considered dataset, 
80 simulations (${\sim}13\%$ of the sample) resulted in prompt collapse 
and ${\sim}40\%$ of the total data is composed by
equal-mass non-spinning binaries.
We include all available resolutions for every binary configuration 
and we treat each point as an independent measure in order to 
improve the characterization of NR uncertainties.
The quasiuniversal relations presented in this work
extend those in Ref.~\cite{Breschi:2019srl} including effects of 
large mass ratios, i.e. $q>1.5$,
and aligned spins with $|\chieff| \lesssim 0.2$.
In App.~\ref{app:eos_qur}, 
we present a recalibration of the quasiuniversal relations
between the PM peak frequency $f_2$ and the NS radius that is not used in {\model} but often employed in GW inference.

The mapping between binary and {\model} parameters is performed on the {\it mass-rescaled} PM parameters using a factorized fitting function (for any quantity $Q$),
\be
\label{eq:pmfitfunc}
Q^{\rm fit} = a_0 \, \, Q^{\rm M}(X) \, \, Q^{\rm S}(\hat S,X) \,\,  Q^{\rm T}(\kt, X) \,,
\ee 
where
$Q^{\rm M} = 1 + a_1^{\rm M} X$ includes the mass ratio contributions in terms of the $X=1-4\nu$ parameter;
$Q^{\rm S} = 1 + p_1^{\rm S} \hat S$ takes into account spin corrections in terms of the spin parameter \cite{Nagar:2019wds}
\be
\label{eq:hatS}
\hat S  = \left(\frac{m_1}{M}\right)^2 \chi_1 + \left(\frac{m_2}{M}\right)^2  \chi_2\,,
\ee
and $p_1^{\rm S} = a_1^{\rm S} (1+ b_1^{\rm S} X)$. The term 
\be
\label{eq:Qtidal}
Q^{\rm T}  =\frac{1+p_1^{\rm T} \,\kt + p_2^{\rm T}\, {\kt}^2}{1+p_3^{\rm T}\, \kt + p_4^{\rm T} \,{\kt}^2}\,,
\ee
takes into account tidal effects in terms of $\kt$ and with $p_i^{\rm T} = a_i^{\rm T} (1+ b_i^{\rm T} X)$.
The coefficients $\{a_i^\cdot, b_i^\cdot\}$ 
are determined fitting the NR data. 
We note that the choice of the fitting function in Eq.~\eqref{eq:pmfitfunc} might be not unique nor optimal; we have experimented with few functions and found Eq.~\eqref{eq:pmfitfunc} sufficiently simple, general and accurate for our purposes. The choice of a rational function for $Q^{\rm T}(\kt)$ is instead motivated by previous works \cite{Bernuzzi:2014kca,Bernuzzi:2015rla,Zappa:2019ntl,Breschi:2019srl}. Finally, we stress the importance of using mass-rescaled quantities in quasiuniversal relations \cite{Bernuzzi:2015rla,Breschi:2019srl}; App.~\ref{app:eos_qur} demonstrates that factorizing the (trivial) binary mass scale is key to obtain EOS-insensitive relations.

The fitting is performed using a least squared method.
Denoting by $Q^{\rm NR}_i$ any NR quantity of interest extracted from the $i$-th
NR simulation and $Q^{\rm fit}_i$ its fit, we define the relative residual of the $i$-th NR simulation,
\be
\label{eq:relres}
r_i = \frac{Q^{\rm fit}_i- Q^{\rm NR}_i}{ Q^{\rm fit}_i}\,,
\ee
and minimize $\chi^2 = \sum_i r_i^2$.
For each calibrated PM parameter,
Table~\ref{tab:fits} reports the calibrated coefficients 
and the associated relative error,
defined as the standard deviation of the relative residuals,
i.e. $\sqrt{{\rm Var}(r_i)}$.
For later purposes (see Sec.~\ref{sec:recal}), we report in Table~\ref{tab:fits} the Kullback–Leibler divergence $\dkl$ 
between the distribution of the residuals $r_i$ and a normal distribution with zero mean and variance ${\rm Var}(r_i)$.
This quantity allows us to verify the Gaussian character of the residuals.

In the following,
we discuss the fit results, i.e. empirical relations for the 
merger properties (Sec.~\ref{sec:fit_mrg}),
for the characteristic PM frequencies and amplitudes 
(respectively Sec.~\ref{sec:fit_fpm} and Sec.~\ref{sec:fit_apm}) 
and for the late-time properties (Sec.~\ref{sec:fit_late}).

 \begin{figure}[t]
	\centering 
	\includegraphics[width=0.49\textwidth]{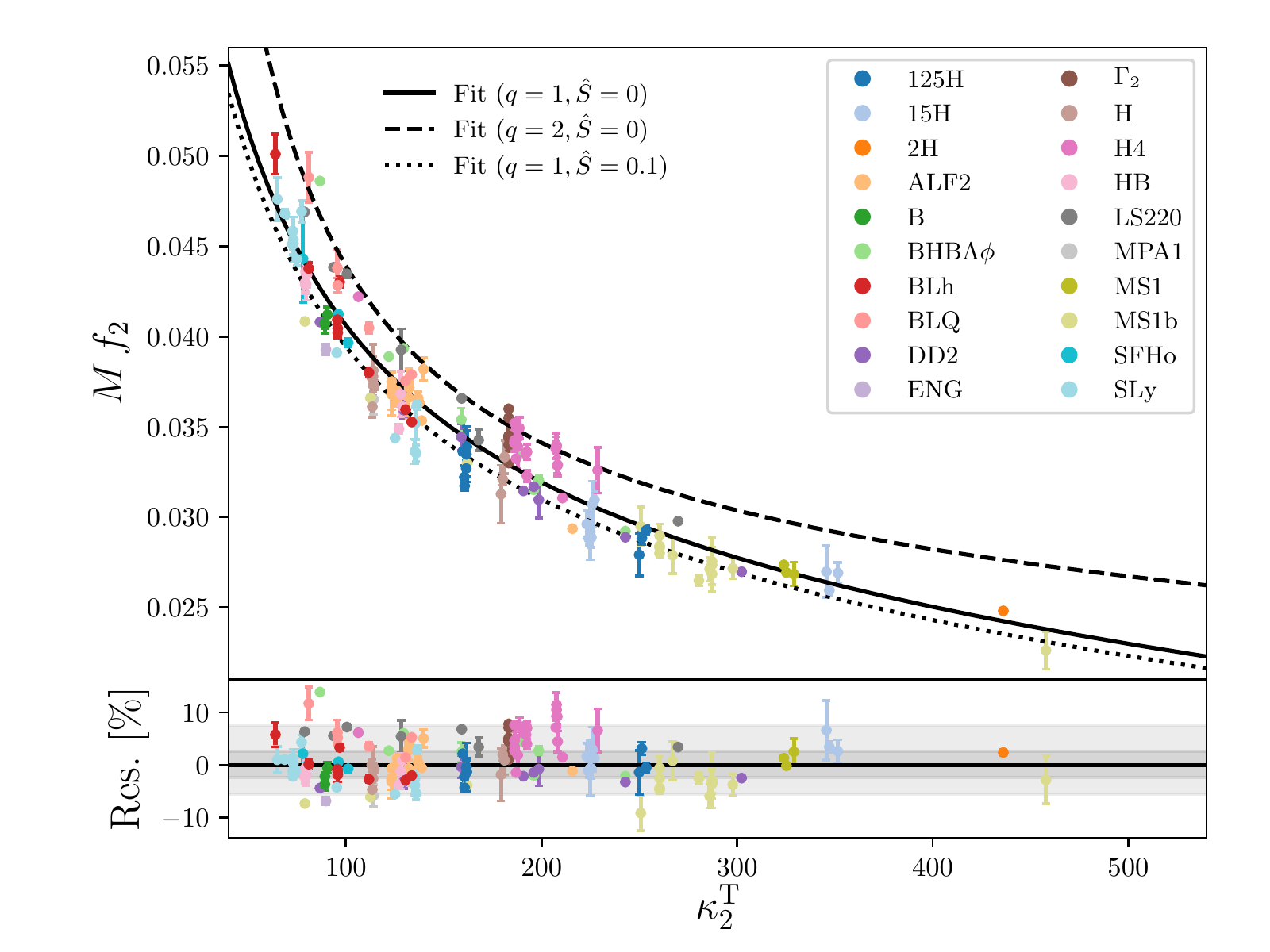}
	\caption{Quasi-universal relation 
					for the PM peak frequency $f_2$
					as function of the tidal polarizability $\kt$.
					Top panel: calibrated relations (black lines)
					compared to NR data (colored dots) 
					extracted from
					the {\core} and the {\sacra} databases.
					Each color corresponds to a different EOS.
					NR medians and error-bars are reported averaging over 
					different numerical resolutions (when available) for the same binary configuration.
					Bottom panel:
					Relative residuals between the calibrated relation and the
					NR data validation set.
					The gray areas show the 
					50\% (dark) and 
					90\% (light)
					credible regions of the residuals.
				 	}
	\label{fig:freqfits}
\end{figure}

 \begin{table*}[t]
   \centering    
   \caption{	Summary of the quasiuniversal relations for the PM parameters $\fitparams$ as functions of the inspiral parameters $\params_{\rm bin}$.
   	The first column report the quantity of interest
   	and the second column shows the range spanned by the available NR data.
	From the third to the fourteenth column, 
	we report the calibrated coefficients of the quasiuniversal relations.
	The last three columns show respectively the $\chi^2$, the relative 
	standard deviation of the fit and the Kullback–Leibler divergence $\dkl$ between 
	the recovered residuals and a normal distribution.}
   \resizebox{\textwidth}{!}{
   \begin{tabular}{cc|cccccccccccc|ccc}        
   	     \hline
     \hline
    $Q^{\rm fit}$ &Range& 
    $a_0$ & $a_1^{\rm M}$ &
    $a_1^{\rm S}$ &$b_1^{\rm S}$ &
    $a_1^{\rm T}$ &$a_2^{\rm T}$ &
    $a_3^{\rm T}$ &$a_4^{\rm T}$ &
    $b_1^{\rm T}$ &$b_2^{\rm T}$ &
	$b_3^{\rm T}$ &$b_4^{\rm T}$ &
     $\chi^2$& Error &$\dkl$\\
     \hline
     \hline
     $ A_{\rm mrg}/M$&  [ 0.159 , 0.313 ]& 
     $0.3948$&$-1.133$&$-0.02992$&$-2.593$&
     $0.03902$&$5.1846\xtt{-5}$& $0.06033$&$1.380\xtt{-4}$&
     $10.41$& $54.51$&$10.83$&$54.54$&
     $0.189$&$1.8\%$&$0.26$\\
     $ A_{0}/M$& [ $2.04\xtt{-4}$ , $0.0699$ ]  &
     $ 0.02356$ & $ 0$ &$ 1.077$ &$ 260.4$ &
     $ -1.318\xtt{-3}$ &$0$ &$0$ &$0$ &
     $ -4.314$ & $0$ & $0$ & $0$ &
     $267$& $66\%$&   $0.14$  \\
     $ A_{1}/M$& [ $0.0262$ , $0.238$]& 
     $-0.05641$&  $-5$&  $-1.135$&  $146.8$&
     $-0.8343$&  $3.882\xtt{-4}$ &  $0.2464$ &  $0$&
     $-5$&  $0$&  $0$&  $0$&
     $13.2$ & $15\%$ & $0.18$   \\
     $ A_{2}/M$&  [ $4.76\xtt{-4}$ ,$0.175$ ] &
     $ 0.1667$ &  $ -5.135 $ &  $ -3.796 $ &  $ -28.47 $ &
    $ 0$ &$ 0$ &  $5.774\xtt{-3}$ &  $ 0$ &
     $0$ & $0$ & $4.027 \xtt{-8}$ &  $ 0$ &
     $78.2$ &$38\% $ & $0.077$ \\
     $A_{3}/M$&[ $5\xtt{-3}$ , $2.04\xtt{-1}$ ] &
     $0.1662$  & $0.1072$  &$-2.046 $  &$-45.06 $  &
     $-7.06 \xtt{-5} $  & $0$  &  $1.354\xtt{-3}$  &$0$  &
     $-1423$  &  $0$  &$284.7$  & $0$  &
     $38.4$  &  $26\%$  &  $0.044$\\
      $M f_{\rm mrg}/\nu$&   [ $0.0554$, $0.141$]&
      $0.2276$  & $0.9233$  & $0.5938 $  & $-1.994 $  &
      $0.03445 $  & $5.58\xtt{-6} $  &  $0.08405 $  &$ 1.133\xtt{-4}$  &
      $13.83 $  & $517.4 $  & $12.75  $  &$139.8 $  &
      $0.431$  &$ 2.6\%$  &$0.45$  \\
      $M f_{2}$& [ $0.0216$ , $0.0512$]& 
      $0.0881$  & $ 22.81 $  &$ 0.2925 $  &$ 25.0 $  &
      $ 0.007023 $  &$ -1.782 \xtt{-6} $  &$0.02587 $  &$ 6.58 \xtt{-6} $  &
      $5.428 $  &$ 0$  &$39.29 $  & $ 0$  &
      $ 0.814$  & $3.9 \% $  & $0.029$     \\
      $M f_{0}$& [ $1.86\xtt{-3}$, $ 0.0441$]& 
      $  0.02734 $  &$   19.32 $  &$   -1.857 $  & $   -75.77 $  &
      $  -2.967 \xtt{-3}$  &$  8.484 \xtt{-6}$  &$  8.584 \xtt{-3}$  &
      $  0$  &$  20.49 $  & $  21.5 $  &$  10.47 $  &$  0$  &
      $  107$  &$   45\% $  &$   0.10$   \\
      $M/ t_{0}$&   [ $8.09\xtt{-3}$, $0.0288$]&
      $0.03265 $  &$0.2994 $  &$ -0.2329 $  &$4.768 $  &
      $ 3.584 \xtt{-3}$  &$0$  &   $ 0.01053 $  &  $ 0$  &
      $ -11.96 $  &  $ 0$  &   $ -3.22 $  & $0.0$  &
      $ 5.11 $  &$9.2 \% $  & $ 0.72 $  \\ 
      $M \,\reblabel{\rm peak}$ & [ $6\xtt{-4}$ ,$6.58\xtt{-3}$ ]&
   $0.1912 $  &$ 4.074 $  &$ -1.573 $  &$ 100$  &
   $ 0.05884 $  &$0$  &  $ 3.896 $  &$0$  &
   $-5.293 $  & $0$  &$0$  &  $0$  &
   $37.7$  & $  27 \%$  &$0.042 $   \\
      $M^2\,\Im(\alpha_{\rm fus})/\nu$& [ $-3.3\xtt{-4}$, $5\xtt{-3}$ ]&
      $  0.003721 $  & 
      $   -1.799 $  &
      $   0.3555 $  &
      $  -7.167 $  &
      $   0.0139 $  &
      $   -2.425 \xtt{-5} $  &
      $   0.05883 $  &
      $    1.882 \xtt{-4}$  &
      $   -28.64 $  &
      $   -36.18 $  &
      $   19.53 $  &
      $   7.089 $  &
      $   346 $  &
      $  75\% $  &
      $   3.7$   \\
      $M\,\Delta_{\rm fm}$& [ $1.5\xtt{-4}$, $0.0423$ ]&
      $0.05139 $  &
      $ 0.4944 $  &
      $ -3.734 $  &
      $-145$  &
      $ -6.25 \xtt{-3}$  &
      $ 1.728\xtt{-5} $  &
      $ 0.01944 $  &
      $ 0$  &
      $-7.936 $  &
      $ 1.882 $  &
      $ 100$  &
      $ 0$  &
      $ 278$  &
      $74 \%$  &
      $0.041 $ \\
      $M\,\Gamma_{\rm fm}$& [ $0$, $0.05$]&
      $  0.1637 $  & $ 209.3$  & $  -0.2997 $  &$  24.5$  &
      $ 0.02195$  &$  0$  &$  0.3528 $  & $  0$  &
      $  -0.5111 $  & $  0.$  &$ 74.72 $  &  $ 0$  &
      $  326$  & $  98 \% $  &$  1.05 $      \\   
     \hline
          \hline
   \end{tabular}}
  \label{tab:fits}
 \end{table*}

 \subsection{Merger properties}
\label{sec:fit_mrg}

Among all the quantities of interest, 
the amplitude and the frequency at merger,
respectively $A_{\rm mrg}$ and $f_{\rm mrg}$,
are properties that can be extracted with good accuracy 
from NR data~\cite{Bernuzzi:2014kca,Breschi:2019srl}.
Our new relations have 1-$\sigma$ errors smaller than $3\%$,
as shown in Tab.~\ref{tab:fits}.
These relations 
are constructed to match the binary black hole (BBH) values for $\kt\to 0 $; the limiting values are taken from the EOB model of Ref.~\cite{Nagar:2019wds}.

The slope parameter $\imalabel{\rm fus}$
characterizes the derivative of 
the GW frequency immediately after merger, 
i.e. $\imalabel{\rm fus} \propto \left({\d f}/{\d t} \right)_{\rm mrg}$.
For every NR simulation,
we estimate this property from the $(2,2)$
time-domain waveform
as the mean value of ${\d f}/{\d t}$
taken in the range $[t_{\rm mrg}, t_0 ]$.
The calibrated relation for $\imalabel{\rm fus}$
shows larger uncertainties compared to $f_{\rm mrg}$,
as reported in Tab.~\ref{tab:fits}.
However, the presented relation shows clear trends
in the tidal parameter and in the mass ratio.
In particular, large-mass-ratio binaries (i.e. $q\gtrsim 1.5$)
show
$\imalabel{\rm fus} \lesssim 0$ due to tidal disruption.

Another early-PM quantity is the time of the 
first amplitude minimum $t_0$.
This quantity is extracted from the time-domain waveform
and can be well captured by the our relations within ${\sim}10\%$.
NR simulations of binaries with $q\gtrsim 1.5$ 
generally show 
increasing $t_0$ due to tidal disruption.
Also the calibrated relations for $ \imalabel{\rm fus} $ and $t_0$ 
include a robust BBH limit for $\kt\to 0$
within the nominal error bars.

\subsection{PM frequencies}
\label{sec:fit_fpm}

We extract the main PM frequency $f_2$ from NR PM spectra 
of the $(2,2)$ mode. 
Generally,
the $f_2$ frequency is estimated as the global maximum of the
PM spectrum; 
however, when modulations are prominent 
and the PM portions are short 
(i.e. $\lesssim 8~{\rm ms}$), 
the $f_2$ contribution is no longer the dominant peak
and it needs to be identified in the local maxima.
As shown in Tab.~\ref{tab:fits},
the quasiuniversal relation for $f_2$ is accurate to 
${\sim} 4\%$ at $1{-}\sigma$ level 
($6{-}7\%$ at $90\%$ credibility level), 
that corresponds to an error of about $100~{\rm Hz}$ ($200~{\rm Hz}$). 
The latter is typically smaller than the FWHM of the spectrum peaks.
Figure~\ref{fig:freqfits} shows this quasuniversal relation: the frequency $Mf_2$ primarily correlates with the tidal polarizability $\kt$,  
while mass ratio and spin contributions
mildly affect the overall value of this quantity.

The bottom panel of Fig.~\ref{fig:freqfits} shows data points with deviations larger than $2{-}\sigma$.
Around $\kt \simeq 207$, it is possible to identify a cluster of NR data corresponding to spinning unequal-mass H4 binaries $1.65{+}1.10~\Msun$ 
with different combinations of
spins~\cite{Dietrich:2016hky,Dietrich:2016lyp}. For these large
mass-ratio cases, the spin correction employed in
Eq.~\eqref{eq:pmfitfunc} will be improved in a future work when more data will be available. 
The largest residual (${\sim}15\%$) is given by the non-spinning equal-mass binaries BHB$\Lambda\phi$ $1.50{+}1.50~\Msun$~\cite{Radice:2016rys}
and BLQ $1.40{+}1.40~\Msun$~\cite{Prakash:2021wpz}.
In both cases,
the remnant collapses into BH shortly after merger, i.e. $t_{\rm coll}\simeq 3~{\rm ms}$,
and the determination of the peak and secondary frequencies from this signal is rather delicate due to the short duration of the transient. 
From the Fourier spectra,
it is possible to identify two dominant broad peaks 
at frequencies 
$M f_{2-0}\simeq 0.036$ and $M f_2\simeq 0.048$ for the BHB$\Lambda\phi$ binary
and
$Mf_{2-0}\simeq 0.036$
and $Mf_{2}\simeq 0.047$
for the BLQ binary.
These values agree with the estimate of $M f_0$ coming from the instantaneous GW frequency; 
however, the peak widths vary depending on the window used to smooth the NR data and it is not possible to clearly identify a carrier frequency and a modulation magnitude from the time-domain waveform.
Consistently with Ref.~\cite{Breschi:2019srl} we chose to identify the second peak with $f_2$ and conservatively include it in the determination of the quasiuniversal relation. In contrast, the choice of the first peak as $f_2$ would be consistent with Ref.~\cite{Radice:2016rys,Prakash:2021wpz}, and the datapoints would not be outliers in the residual plot.

The value of the frequency $f_0$ is estimated as 
$f_0 = \tsfzero^{-1}\simeq (t_3-t_1)^{-1}$ (Sec.~\ref{sec:nodes}).
The frequency $f_0$ shows a non-monotonic 
dependency on the tidal coupling $\kt$ for $q\simeq1$.
The relative error associated with $f_0$ is ${\sim}60\%$,
that is considerably larger than the error on the peak frequency $f_2$.
This uncertainty can be related to the method used to estimate $f_0$
and to the numerical error that affects amplitude fluctuations.
In principle, the frequency $f_0$ can be also extracted from the 
$(\ell=2,m=0)$ mode of the GW waveform. 
However, numerical errors appear to be larger for HM components,
due to the lower magnitude of the strains,
and the corresponding spectra do not show neat and unambiguous 
Fourier peaks, yielding to less accurate calibrated relations. 

\subsection{PM amplitudes}
\label{sec:fit_apm}

The PM amplitudes $A_1$, $A_2$ and $A_3$
are extracted from the time-domain NR data and they
show a decreasing trend for increasing $\kt$
and for increasing mass ratio,
similarly to Ref.~\cite{Breschi:2019srl}.
This can be understood as the effects of 
stiffer EOSs and larger mass ratios that produce 
less violent dynamics in the remnant (for a fixed $M$).
As a consequence of tidal disruption, the first amplitude $A_0$ 
increases with increasing mass ratio. Overall, these quantities show errors between $15\%$ and $40\%$,
except for $A_0$, which shows an error $\gtrsim 60\%$ 
since this quantity is comparable in magnitude to NR errors.

\subsection{Late-time features}
\label{sec:fit_late}

The damping time $\reblabel{\rm peak}$ of the 
decaying tail in {\model} is estimated from NR data 
using the approximation for exponential sinusoidal functions, i.e.
$\reblabel{\rm peak}\simeq  {\rm max}(A(t))/[2\,{\rm max}(A(f))] $,
where ${\rm max}(A(t))$ is the maximum amplitude of the time-domain waveform
and ${\rm max}(A(f))$ is the maximum amplitude of the frequency-domain spectrum.
Despite errors of ${\sim}30\%$, 
the calibrated relation has a physically reasonable trend. For example, $\reblabel{\rm peak}$ decreases for increasing mass ratios, in agreement with the tidally disruptive dynamics of high-mass ratio mergers.

The FM displacement $\Delta_{\rm fm}$ is estimated 
from the time-domain NR waveforms as the largest displacement 
in the instantaneous GW frequency $f_{22}(t)$
from the PM peak $f_2$.
The $\Delta_{\rm fm}$ predictions
show similar trends and comparable values 
to $f_0$ for equal-mass binaries.
More significant differences emerge instead as the mass-ratio get larger.
The FM damping time $\Gamma_{\rm fm}$ is also 
estimated from the time-domain NR data 
fitting a damped sinusoidal to the instantaneous GW frequency. 
This quantity has the less accurate relation
among the presented cases (${\sim}90\%$) due to the large 
errors introduced by the extraction method.

\section{Validation}
\label{sec:valid}

We validate the {\model} model by computing its faithfulness 
$\F$ against 102 NR waveforms of Refs.~\cite{Radice:2017zta,Radice:2018pdn,
				Breschi:2019srl,Nedora:2020pak,
				Bernuzzi:2020txg,
				Camilletti:2022jms}
that were not used for the calibration.
Among the considered simulations,
12 binaries show prompt collapse into BH.
The validation set is composed by 
NR simulations of non-spinning BNS
performed with {\scshape THC}~\cite{Radice:2012cu}
that include different neutrino treatments,
turbolent viscosity schemes
and five EOSs, i.e.
BHB$\Lambda\phi$~\cite{Banik:2014qja},
DD2~\cite{Typel:2009sy}, 
LS220~\cite{Lattimer:1991nc}, 
SFHo~\cite{Steiner:2012xt}
and SLy~\cite{Douchin:2001sv}.
The intrinsic binary properties 
cover the 
ranges $M\in[2.6,3.4]~\Msun$, 
$q\in[1,1.8]$
and $\kt\in[47,199]$.
The unfaithfulness $\Fbar=1-\F$ between two waveform templates, say $h_1$ and $h_2$, is defined as
 \be
 \label{eq:fbar}
 \Fbar(h_1,h_2) = 1 - \max_{t_{\rm mrg},\phi_{\rm mrg}} \frac{(h_1|h_2)}{\sqrt{(h_1|h_1)(h_2|h_2)}}\,,
 \ee
 where the maximization is performed over 
 the coalescence time and phase, respectively 
 $t_{\rm mrg}$ and $\phi_{\rm mrg}$.
 The inner product $(h_1|h_2)$ is
 \be
  \label{eq:innerprod}
 (h_1|h_2) = 4 \Re \int \frac{h_1^*(f) \,h_2(f)}{S_n(f)}\, \d f\,,
 \ee
 where $S_n(f)$ is the power spectral density (PSD)
 of the detector.
We employ the PSD curve of the next-generation detector 
ET~\cite{Hild:2010id,Hild:2011np}
(configuration D).
The unfaithfulness is computed between 
the PM part of the NR waveform
and the {\model} for the same intrinsic parameters,
i.e. $\Fbar(h_{\rm NR}, h_\text{\model})$, over 
the frequency range $[1, 8]~{\rm kHz}$.

Moreover, we compare the NR faithfulness of {\model} to that of the 
time-domain {\oldmodel} model introduced 
in Ref.~\cite{Breschi:2019srl}. Note that {\oldmodel} can be also enhanced with the 
parameters $\{\alpha,\beta,\phi_{\rm PM}\}$ \cite{Breschi:2021xrx}, that are analogous to $\{\adrift,t_{\rm coll},\phi_{\rm PM}\}$ 
for ${\model}$  
and further discussed in App.~\ref{app:nrpm_newpars}.
The main differences between the two models
are the following.
The frequency evolution of {\model} around merger 
is fully calibrated on NR data, 
while {\oldmodel} uses a post-Newtonian approximation.
The quasiuniversal relations used in {\oldmodel} are not calibrated on {\sacra} data, although they are compatible with the new ones computed here for ${\model}$.
Moreover, 
{\model} includes a full description of damped 
FM effects and it permits the calibration of the collapse time $t_{\rm coll}$, that improves the characterization
of the $f_2$ peak.

In the following sections, 
we discuss the introduction of the recalibration parameters for both time- and frequency-domain models (Sec.~\ref{sec:recal})
and we present the unfaithfulness results 
(Sec.~\ref{sec:fbar}) computed on the independent
validation set of NR data.

\subsection{Recalibrations}
\label{sec:recal}

 \begin{figure*}[t]
	\centering 
	\includegraphics[width=0.49\textwidth]{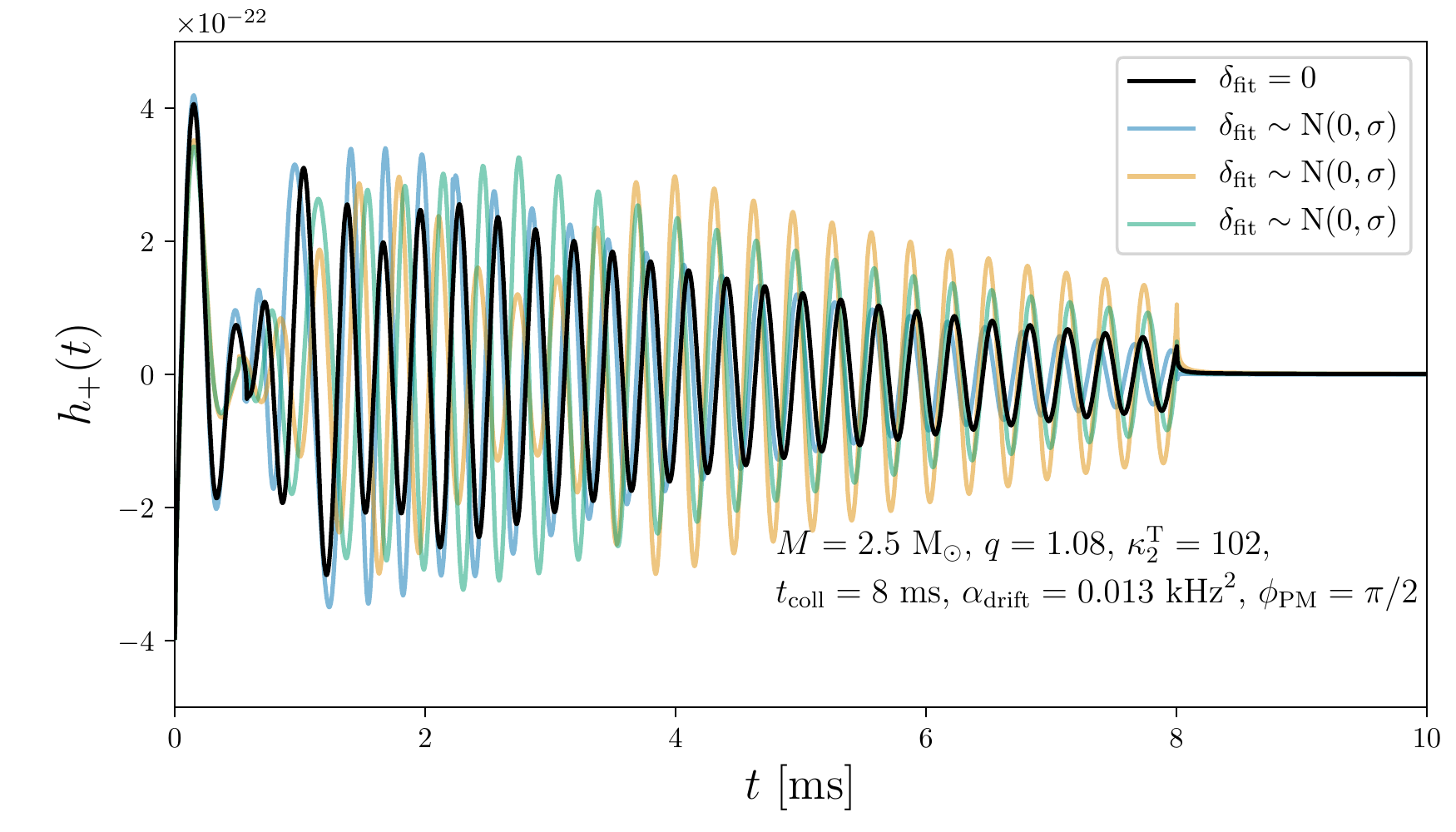}
	\includegraphics[width=0.49\textwidth]{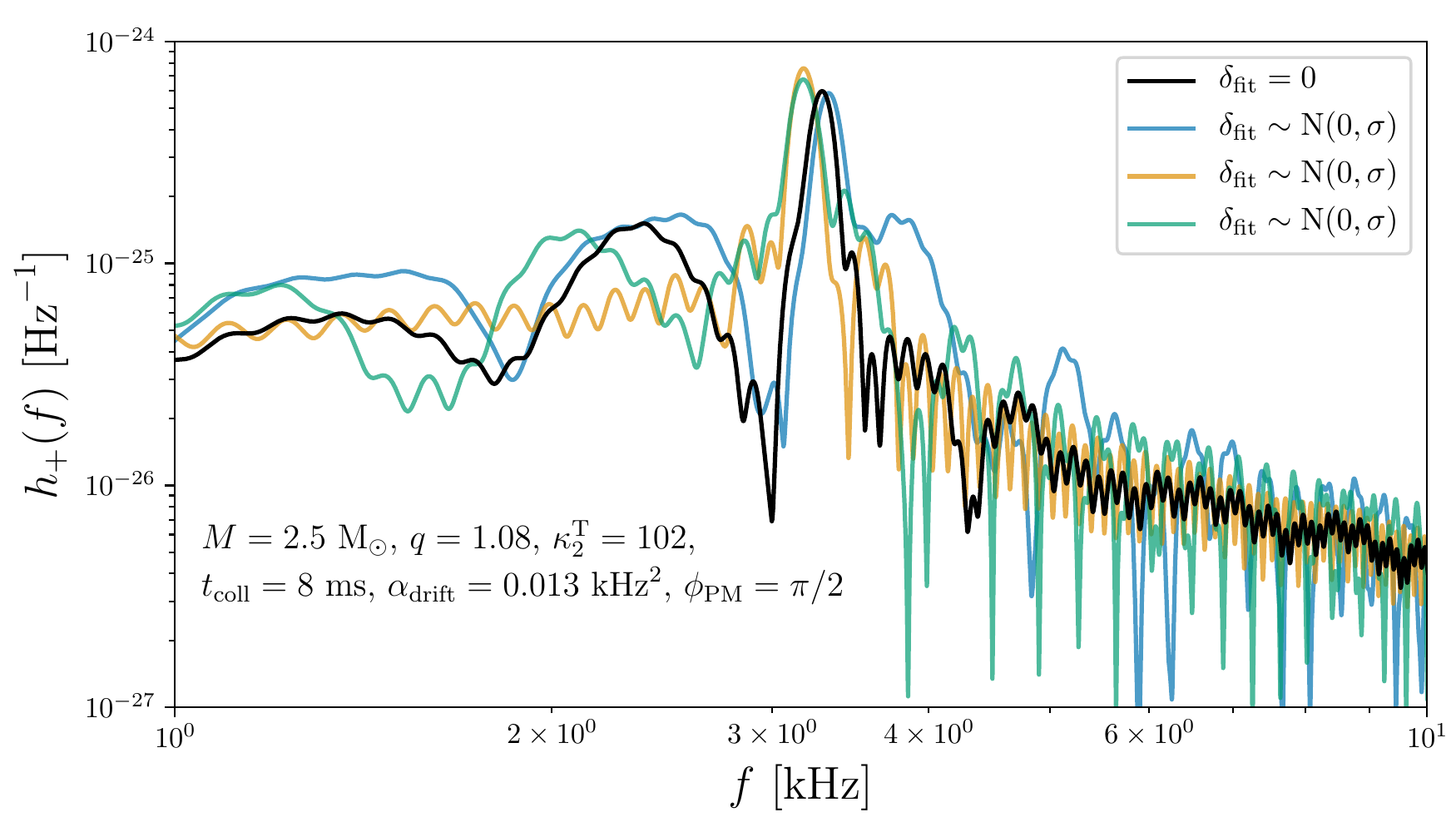}
	\caption{Effect of recalibration terms on {\model} waveform.
					The figures show exemplary templates 
					of the GW plus polarization $h_+$ 
					in the time-domain (left) and in the frequency-domain 
					(right).
					The template has been
					computed for the parameters $M=2.5~\Msun$,
					$q=1.08$, $\kt=102$, 
					$t_{\rm coll}=8~{\rm ms}$,
					$\adrift=0.013~{\rm kHz}^2$,
					$\phi_{\rm PM}=\pi/2$
					and
					locating the source at a luminosity distance of $40~{\rm Mpc}$.
					Black lines show the exact {\model} predictions,
					i.e. the recalibration parameters 
					are identically zero, $\recalibpm=0$.
					The colored lines show three exemplary cases
					where the values of the recalibrations $\recalibpm$
					have been randomly extracted from a zero-mean 
					normal distribution with variance prescribed by the 
					errors of the residuals.}
	\label{fig:recalwf}
\end{figure*}

 The EOS-insensitive relations
 developed in Sec.~\ref{sec:calib} 
carry intrinsic uncertainties due to small violations of universality (EOS dependence) and/or fitting inaccuracies. 
Calibration errors of the empirical relations should be taken into account every time such mappings are employed, 
in particular during the calculation of fitting factors and during parameter estimation, in order to perform robust predictions.
This can be done by introducing appropriate parameters associated with the fluctuation of the residuals.
A by-product of this process is that the model can improve its performance in describing the data.

Labeling $Q$ a generic quantity 
estimated from a quasiuniversal relation
calibrated on NR data,
we introduce an associated {\it recalibration} $\delta_Q$ that affects the prediction
$Q^{\rm fit}$
of the EOS-insensitive relation as 
\be
\label{eq:recal-effect}
 Q= Q^{\rm fit}\,(1+\delta_Q)\,.
 \ee
The recalibration $\delta_Q$
corresponds to 
a fractional displacement from the 
prediction $Q^{\rm fit}$ 
of the quasiuniversal relation.
The recalibration procedure employed here is similar to the 
spectral calibration envelopes used in GW analyses~\cite{Vitale:2011wu}.
However, here we aim to integrate the model's uncertainties 
in the inference rather than the instrumental errors.
A similar approach has been used in \cite{Breschi:2021tbm} (Sec.~5).

In GW inference applications, 
the recalibrations of each calibrated PM property
are treated as standard parameters.
In this context, it is key that the prior
 distribution used in the inference 
 is a good representation of the residuals
 of the EOS-insentive relation.
This allows us to perform a rigorous marginalization on
the theoretical uncertainties of the model,
delivering more robust and conservative estimates.
Interestingly, under the assumption that the NR error is subdominant compared to the physical breaking of quasiuniversality, 
the measurement of the recalibration parameters from the data could also be used to distinguish 
between different EOSs
and observatively probe the breaking of quasiuniversality.

A robust characterization
of the NR errors is needed in order to employ
a coherent prior distribution for the 
recalibration parameters in the GW inference
routines.
In principle,
the uncertainties associated to an EOS-insensitive
relation can be estimated 
as functions of the employed parameters using regressive methods
or parameter estimation techniques.
Following the methods of Ref.~\cite{Vitale:2011wu,Breschi:2021tbm},
an alternative and simpler approach
is to consider a normally distributed 
prior distribution 
with variance prescribed by the 
errors of the residuals (see Tab.~\ref{tab:fits}).
Thus, the relative errors of the EOS-insensitive relations 
are key quantities, since they define the theoretical uncertainties of the model.

Figure~\ref{fig:recalwf} illustrates the use of recalibrations 
in {\model} for an examplary case.
The recalibration
parameters $ \recalibpm=\{\delta_i\}$
are considered for each element of $\fitparams$.
These additional degrees of freedom 
mildly affect the merger portion, i.e. $t<t_0$
due to the accuracy of the empirical relations close to merger.
However,
the recalibration coefficients 
have a larger effect on the late-time PM features
whose EOS-insensitive relations 
introduce larger uncertainties.
Analogously, the recalibrations can be introduced 
for {\oldmodel}.
This additional flexibility is expected to
significantly improve the data fitting 
by adjusting the PM morphology of the template
to match the targeted signal,
similarly to agnostic approaches~\citep[e.g.][]{Chatziioannou:2017ixj,Easter:2020ifj}.

\subsection{Unfaithfulness}
\label{sec:fbar}

 \begin{figure}[t]
	\centering 
	\includegraphics[width=0.49\textwidth]{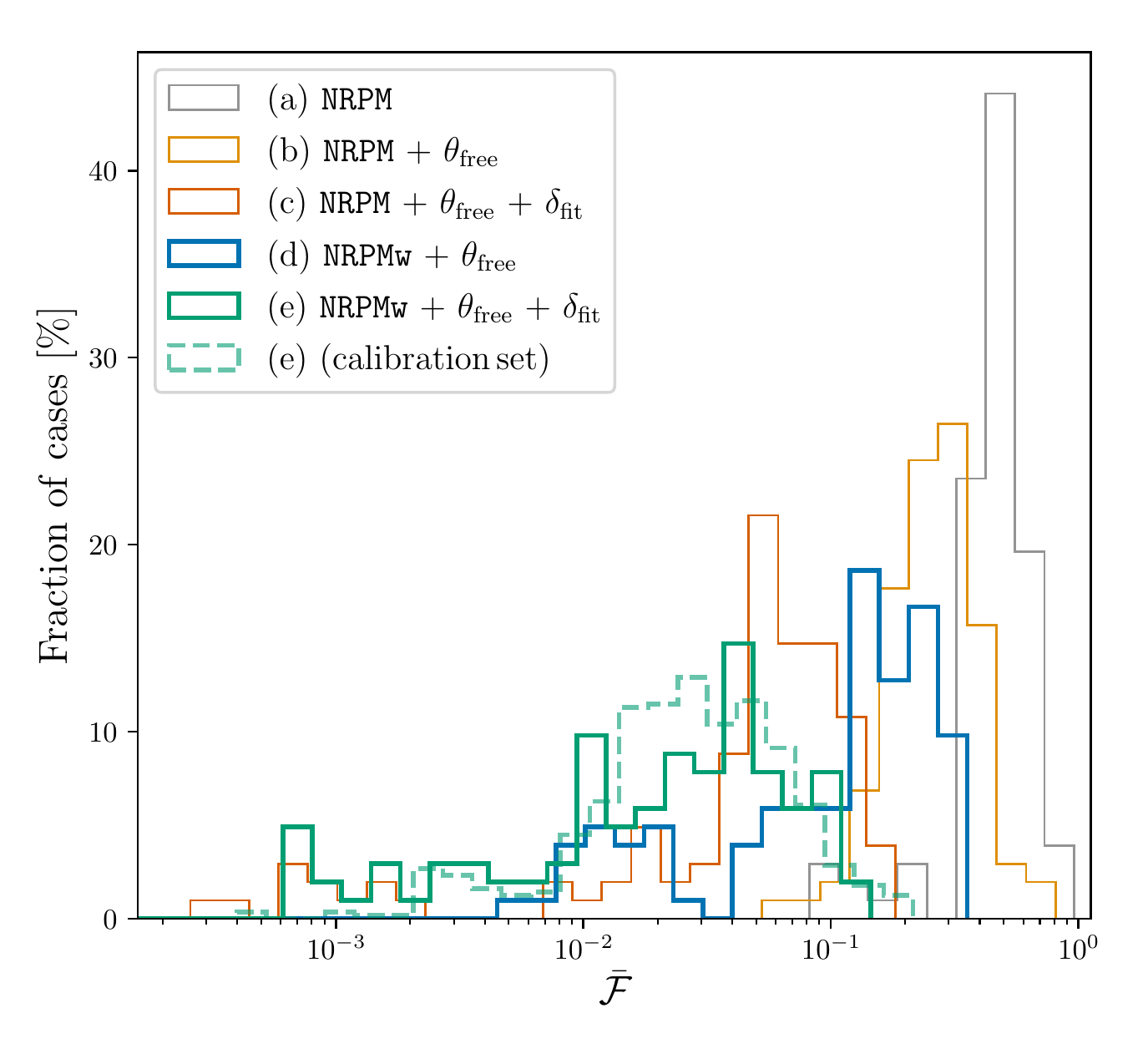}
	\caption{Recovered unfaithfulness $\Fbar$ 
		between PM models and NR data of the 
		validation set~\cite{Radice:2017zta,Radice:2018pdn,
			Breschi:2019srl,Nedora:2020pak,
			Bernuzzi:2020txg,
			Camilletti:2022jms}
		employing ET-D sensitivity~\cite{Hild:2010id,Hild:2011np}.
		For {\oldmodel}~\cite{Breschi:2019srl}
		(thin lines),
		we compute $\Fbar$ with the standard model (a),
		including PM parameters (b)
		and also the recalibrations (c).
		Analogously,
		the $\Fbar$	recovered for {\model}
		(thick lines) include
		the PM parameters (d) and also the recalibrations (e).
		The dashed histogram shows
		the $\Fbar$ for case (e) computed over 
		the calibration set.
	}
	\label{fig:match}
\end{figure}

\begin{figure*}[t]
	\centering 
	\includegraphics[width=0.99\textwidth]{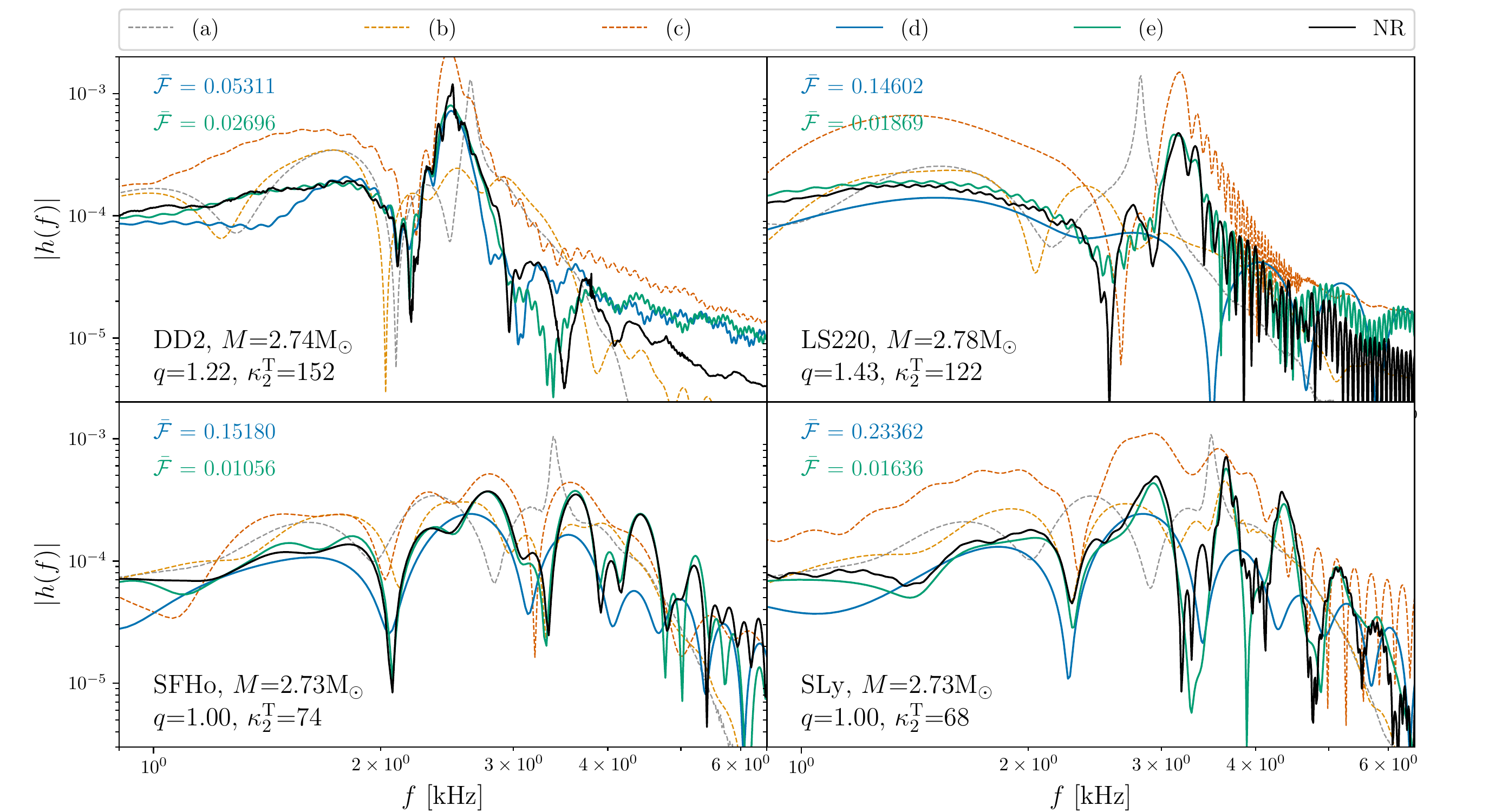}
	\caption{Comparison
		between PM models and 
		exemplary NR data of the validation set.
		Colored lines show the spectra for the different models,
		analogously to Figure~\ref{fig:match}.
		Solid lines are used for {\model} spectra
		and dashed lines are employed for {\oldmodel}.
		NR spectra are reported with black solid lines.
		The plot includes also the corresponding unfaithfulnesses estimated with {\model} model,
		i.e. case (d) in blue and case (e) in green.
	}
	\label{fig:compare}
\end{figure*}

We compare here the NR faithfulness results for {\model} and {\oldmodel}.
In Figure~\ref{fig:match}, we report histograms of the unfaithfulness computed on the validation NR sample of
\begin{itemize}
\item[(a)] {\oldmodel} without resorting to minimization methods;
\item[(b)] {\oldmodel} minimizing over the additional PM parameters $\{\alpha,\beta,\phi_{\rm PM}\}$
and setting $\recalibpm=0$;
\item[(c)] {\oldmodel} with recalibration parameters $\recalibpm$ and minimizing over $\recalibpm$ and $\{\alpha,\beta,\phi_{\rm PM}\}$;
\item[(d)] {\model} minimizing over the additional PM parameters $\freeparams$
and setting $\recalibpm=0$;;
\item[(e)] {\model} with recalibration parameters $\recalibpm$ and minimizing over $\recalibpm$ and $\freeparams$.
\end{itemize}
In particular, the minimization procedure 
is performed as follows.
For each NR waveform,
we compute the corresponding {\model} 
(or {\oldmodel}) template
fixing the intrinsic parameters $\params_{\rm bin}$
to the values of the NR simulation
and estimating the additional parameters
($\freeparams$ and $\recalibpm$)
minimizing the unfaithfulness $\Fbar$, 
i.e. Eq.~\eqref{eq:fbar},
using a differential evolution method~\cite{Storne:1997}.
For each case and for each NR data,
the additional degrees of freedom are independently 
varied over a physically-motivated range
\footnote{For {\model},
					we set the time of collapse $t_{\rm coll}\ge t_0$,
					 the frequency drift $M^2\adrift \in [-10^{-5}, 10^{-5}]$,
					 the PM phase $\phi_{\rm PM} \in [0, 2\pi]$
					 and the recalibrations $\delta_i \in [-4 \sigma_i, +4 \sigma_i]$,
					 where $i$ runs over the calibrated PM 
					 quantities and $\sigma_i$ is the corresponding
					 standard deviation of the NR residuals
					 (see Tab.~\ref{tab:fits}).
				 	}
in order to estimate the minimum $\Fbar$.

Case (a) gives results comparable to \cite{Breschi:2019srl}, with median value $\Fbar$ equal to $0.45$ 
and few cases with $\Fbar\le0.1$ ($2\%$).
Indeed, the only differences between this work and Ref.~\cite{Breschi:2019srl} are
the PSD and the different validation set.
In case (b), the inclusion of the free parameters $\{\alpha,\beta,\phi_{\rm PM}\}$ 
improves the faithfulness of {\oldmodel} 
by shifting the median value to $\Fbar\simeq 0.27$,
but the majority of the recovered values ($97\%$) 
lies above $\Fbar=0.1$.
The additional inclusion of the recalibration parameters,
shown in case (c),
considerably enhances the quality of the recovered waveforms, since $\Fbar$ decreases 
with median $\Fbar={0.06}$
and down to $\Fbar\sim O(10^{-3})$,
corresponding to short-lived remnants
and prompt BH collapses.
The fraction of cases with $\Fbar<0.1$
corresponds to $83\%$ and we recovered
$\Fbar<0.2$ for all binaries in the validation set.

Moving to the novel {\model} model,
case (d) show an overall improvement in the faithfulness
compared to the equivalent case (b), 
with median $\Fbar{\sim}0.13$
and a fraction of $38\%$ with $\Fbar<0.1$. 
We attributed this enhancement to the modeling choices employed in {\model}, 
since the number of parameters minimized ($\freeparams$) 
is the same as case (b).
Moreover, case (d) shows a small cluster with 
$\Fbar\lesssim 3{\times}10^{-2}$ (${\sim}20\%$), 
mainly populated by short-lived 
remnant and prompt BH collapses.
In case (e), the additional inclusion of recalibration terms considerably improves 
the agreement of {\model} to the NR data. We obtain a median $\Fbar$ of $2.5{\times}10^{-2}$
and report $94\%$ of the validation set with $\Fbar< 0.1$.
We recover similar statistics 
applying case (e) over the six-hundred 
NR simulations of the calibration set,
shown with dashed line in Figure~\ref{fig:match}.
Moreover,
the histogram (e) shows that the cluster constituted 
by short-duration signals moves toward $\Fbar=10^{-2}$
and we recovered values comparable to or smaller than $\Fbar=3{\times}10^{-2}$ for several long-duration transients,
such as SLy $1.30{+}1.30~\Msun$,
and unequal-mass binaries,
such as DD2 $1.50{+}1.25~\Msun$.
The overall improvement with respect to 
the comparable case (c) is roughly half order of magnitude.

The recovered results validate the modeling choices,
suggesting that the primary contributions of the theoretical errors are
the inaccurate predictions of the EOS-insensitive relations.
Considering the faithfulness condition proposed in
	Ref.~\cite{Damour:2010zb,
		Chatziioannou:2017tdw,Gamba:2020wgg}
	and fixing $N=9$ as number of intrinsic parameters 
	$\{\params_{\rm bin},\freeparams\}$,
	the recovered upper-bound accuracy $\Fbar \simeq 10^{-1}$
	of {\model} in case (e) can be translated 
	into a model robustness threshold of SNR ${\sim}7$.
	Above this threshold, systematic waveform errors can become relevant.
	The threshold moves to SNR ${\sim}11$ if we include 
	the recalibrations $\recalibpm$ as intrinsic parameters,
	i.e. $N=22$.
	On the other hand,
	employing the recovered median value $\Fbar \simeq 2.5{\times}10^{-2}$,
	we estimate a faithfulness threshold SNR equal to $13$ 
	for $N=9$ and $21$ for $N=22$.
	Considering an averaged threshold of SNR ${\sim} 10$,
	this limit matches the requirements imposed by 
	ET detector for
	(optimally-oriented) sources located 
	at luminosity distances  $\gtrsim 40~{\rm Mpc}$~\cite{Tsang:2019esi,Breschi:2019srl,Easter:2020ifj,
		Breschi:2021wzr,Breschi:2021xrx}.

Notably, 
the $\Fbar$ values computed on simulations
with different grid resolution or physical schemes 
suffer from considerable fluctuations for some binaries.
Some examples are:
LS220 $1.47{+}1.27~\Msun$
that gives $\log_{10}\Fbar=-0.84$
at standard resolution without turbolent viscosity 
and $\log_{10}\Fbar=-1.42$
at high resolution with turbolent viscosity;
and LS220 $1.35{+}1.35~\Msun$ (with turbolent viscosity)
that gives $\log_{10}\Fbar=-1.05$
at standard resolution 
and $\log_{10}\Fbar=-1.79$
at low resolution.
These results suggest that the largest $\Fbar$ might be related 
to an inaccurate modeling of the late-time features
or to an excess of numerical error in the data.
On the other hand,
the accuracy of NR templates computed
from different grid resolutions spans the range $\Fbar\simeq 0.6$
to $\Fbar\simeq 10^{-2}$~\cite{Breschi:2019srl},
comparably to case (c) and (e). 
These non-negligible errors originate from 
finite resolution of numerical data.

Figure~\ref{fig:compare} shows the comparison between the PM model spectra
and NR data for four exemplary cases
extracted from the validation set.
The first case is DD2 $1.509{+}1.235~\Msun$ 
which generates a long-lived remnant, $t_{\rm coll} \sim O(100~{\rm ms})$.
The {\oldmodel} model (a) predicts an erroneous $f_2$ peak,
which biases the estimation of the damping time in case (b).
The result improves to $\log_{10}\Fbar=-1.3$ in case (c).
The novel {\model} matches well the NR data, delivering  
$\log_{10}\Fbar=-1.3$ in case (d) 
and $\log_{10}\Fbar=-1.6$ in case (e).
The second case is LS220 $1.635{+}1.146~\Msun$ 
with tidal-disruptive behavior that collapses into BH 
${\sim}12~{\rm ms}$ after merger.
For this simulation, {\model} (d) is not capable to match the 
dominant PM peak returning $\Fbar\simeq 0.15$. 
Then, the recalibrations (e) strongly improve
the agreement to NR data, yielding $\log_{10}\Fbar=-1.7$.
The third case is SFHo $1.364{+}1.364~\Msun$
which generates a short-lived remnant with 
$t_{\rm coll}\simeq 4~{\rm ms}$.
This spectrum highlights the relevance of modulation effects
in PM signals. The comparison shows the flexibility of the
recalibrated {\model} (e) in capturing the several
Fourier peaks, delivering $\log_{10}\Fbar=-1.9$.
The last case is SLy $1.364{+}1.364~\Msun$ 
with $t_{\rm coll}\simeq 12~{\rm ms}$
that shows prominent modulations in the spectrum.
{\oldmodel} does not match well the prominent
subdominant peaks returning $\log_{10}\Fbar=-0.7$ in case (c).
This result is similar to {\model} (d) but 
considerably improved 
with the inclusion of recalibrations (e) to $\log_{10}\Fbar=-1.8$.

\section{Conclusions}
\label{sec:conclusion}

This paper presents {\model}, a frequency-domain model for PM GW from BNS remnants calibrated with EOS-insensitive relations from the largest publicly-available set of NR simulations. {\model} is designed to be employed in fully or partially informed Bayesian inference from GW data. {\model} includes the dependency on the intrinsic binary parameters through the EOS-insensitive relations, thus allowing (i) the direct astrophysical inference of \textit{all} the BNS parameters without assuming a pre-merger signal/detection, and at the same time (ii) a phase-coherent attachment with pre-merger templates \cite{Breschi:2019srl}. The current uncertainties of EOS-insensitive relations can be taken into account in a partially informed approach using recalibrations parameters. This enhances the flexibility of the model in capturing the complex morphology of PM signals and improves fitting factors. We stress that a recalibration procedure similar to the one introduce here should be employed every time EOS-insensitive are applied to any type of data.

{\model} was validated with an independent set of 102 NR simulations. The fitting factors favorably compares against the results obtained with similar frequency-domain models~\cite{Tsang:2019esi} and significantly improve those we obtained with {\oldmodel}~\cite{Breschi:2019srl}. The improvement is mainly related to a more accurate modeling of the {\it merger} features and to an improved description of the FMs when compared to {\oldmodel}. The faithfulness of the recalibrated {\model} is comparable to that obtained using unmodeled (non-informed) templates and agnostic approaches~\cite{Easter:2020ifj,Soultanis:2021oia}. This comes at the cost of 13 recalibration parameters and three free parameters, compared to the typical $O(10)$ parameters of unmodeled templates. However, differently from the latter, {\model} delivers complete posteriors for the BNS parameters, including mass, mass ratio, etc. The {\model} faithfulnesses are comparable to the accuracy of current NR templates for BNS remnants. The further development of high-precision NR simulations is key for the design of robust PM models.

{\model} builds on a new set of EOS-insensitive relations for the PM spectra. We have focused the development of quasiuniversal relation that employ the tidal coupling constant $\kt$ in view of utilizing the model as an EOB completion \cite{Bernuzzi:2014kca,Bernuzzi:2015rla,Zappa:2019ntl,Breschi:2019srl}. 
The most robust relations we obtained are, not surprisingly, the merger amplitude, the merger frequency and the dominant $f_2$ peak. The $1{-}\sigma$ uncertainties of these relations are of the order of {4\%} due to either uncertainties of NR data or EOS-dependent features.

We found that the presence of softening effects due to quark deconfinement or hyperonic degrees of freedom at high densities does {\it not} introduce significant deviations
above the $2{-}\sigma$ credibility level
in the EOS-insensitive relation for $f_2$ developed here. 
Hence, the observational imprint of EOS softening might be better revealed from an earlier BH collapse phenomenology, e.g.~\cite{Sekiguchi:2011mc,Radice:2016rys,Prakash:2021wpz,Huang:2022mqp,
Fujimoto:2022xhv}, rather than from the measurement of PM frequencies (under the assumption that our sample of models adequately represent the ``true'' EOS). 
However, 
the cases BHB$\Lambda\phi$ $1.50{+}1.50~\Msun$, 
BLQ $1.40{+}1.40~\Msun$~\footnote{As discussed in Sec.~\ref{sec:fit_fpm}, 
	these massive binaries of Ref.~\cite{Radice:2016rys,Breschi:2019srl,Prakash:2021wpz} are marginal cases very close to prompt collapse and its interpretation is not fully clear with the present data.}
and other literature results~\cite{Bauswein:2018bma,
	Chatziioannou:2019yko,
	Most:2019onn}
suggest that the EOS-insentive relations for $Mf_2(\kappa_2^{\rm T})$ (and in principle for other quantities) 
might break for particular binary masses and in presence of ``strong'' phase transitions~\footnote{
The term ``strong'' is often used in the literature but it does not have any precise meaning; in this context it is used as a tautology to indicate that the EOS model breaks the quasiuniversal relation.
}.
This opens the possibility of using {\model} to identify this new extreme matter physics via Bayesian analyses following the method of~\cite{Breschi:2019srl,Wijngaarden:2022sah}.
We stress that these types of analyses strictly probe only the violation of the particular quasiuniversal relation that is assumed in a model. Hence, the robust construction of EOS-insensitive relations and the use of recalibration parameters are key for the interpretation of the inference results. Future analysis must incorporate the here proposed recalibration parameters in the inference, because assessing the breaking of the quasiuniversality requires the knowledge of the theoretical uncertainty of the EOS-insentive relation.

The use of dimensionless and mass-rescaled quantities is a key aspect in building EOS-insensitive relations. An example illustrating this fact is the breaking of the quasiuniversal relations claimed in Ref.~\cite{Raithel:2022orm}. The latter refers to relations of type $f_2(R)$ that are different from those employed here. In Appendix~\ref{app:eos_qur}, we verified that those $f_2(R)$ relations are broken also by some data of the {\core} database. The additional term proposed in \citet{Raithel:2022orm} does not fix the breaking of some {\core} data with softening effects at high densities. However, we verified that the use of mass-rescaled quantities, i.e. $Mf_2(R/M)$, leads to more robust EOS-insensitive relations. Hence, considering $Mf_2(R/M)$ or $f_2(R)$ in a Bayesian analysis of the same data would incorrectly lead to two different conclusions about the EOS. 
In a companion paper we report a study on the application of {\model} to detection and Bayesian parameter estimation of PM signal with ET.
As anticipated by the faithfulness calculations presented here, {\model} can improve the performances of {\oldmodel}~\cite{Breschi:2019srl},
yielding to threshold SNRs
comparable to those of unmodeled analyses~\cite{Chatziioannou:2017ixj,Easter:2020ifj}.
For example, the model can be used to infer 
the dynamical frequency evolution of the remnant
and the time of BH collapse already at the minumum SNR treshold.
Under the important caveat on the robusteness of the assumed EOS-insensitive relations discussed above, these observables can provide insight into the properties of matter under extreme conditions. 
Moreover, {\model} can be employed 
together with inspiral-merger templates 
to characterize the full GW spectrum of 
BNS mergers following~\cite{Breschi:2019srl,Breschi:2021xrx}. Further studies in this directions will be reported in a third paper of this series.

\section*{Acknowledgments}

MB, SB and KC acknowledge support by the EU H2020 under ERC Starting Grant, no.~BinGraSp-714626. MB acknowledges support from the Deutsche Forschungsgemeinschaft (DFG) under Grant No. 406116891 within the Research Training Group RTG 2522/1. 
SB acknowledges the hospitality of KITP at UCSB and partial support by the National Science Foundation under Grant No. NSF PHY-1748958 during the conclusion of this work.
AC and AP acknowledge PRACE for awarding them access to Joliot-Curie at GENCI@CEA. They also 
acknowledge the usage of computer resources under a CINECA-INFN agreement (allocation INF20\_teongrav and INF21\_teongrav).
The computations were performed on {\scshape ARA}, a resource of Friedrich-Schiller-Universt\"at Jena supported in part by DFG grants INST 275/334-1 FUGG, INST 275/363-1 FUGG and EU H2020 BinGraSp-714626.
and on the {\scshape Tullio} sever at INFN Turin.

The waveform model developed in this work, {\model}, is implemented in {\scshape bajes} and the software is publicly available at:

\url{https://github.com/matteobreschi/bajes}.

\appendix

\section{Wavelet approximations}
\label{app:numerror}

We discuss the approximations employed to
compute the frequency-domain wavelet $\wavelet(f)$
Eq.~\eqref{eq:fdwavelet} for different 
values of the $\alpha$ parameter.

When $\alpha$ is identically zero, 
Eq.~\eqref{eq:tdwavelet} reduces to a damped sinusoidal function,
but Eq.~\eqref{eq:fdwavelet} leads to an indeterminate form.
Then, the latter can be replaced by 
\be
\label{eq:wave_damped}
\wavelet(f) = \e^\gamma\,\left(\frac{\e^{\tau\zeta}-1}{\zeta}\right)\,,
\ee
where $\zeta(f) = \beta - 2\pi\i f $.
Eq.~\eqref{eq:wave_damped} represents a good approximation
also when the wavelet is strongly damped, i.e. $\reb$ dominates over
the quadratic contributions.
This is crucial to simplify the computations for higher-order FM 
terms, whose damping time decrease linearly with the 
approximation order (see App.~\eqref{app:fmapprox}).

For $|\alpha|\ll 1 $, 
arithmetic overflows arise 
in numerical computations.
Then, for these cases,
we expand Eq.~\eqref{eq:tdwavelet}
 around small values of $\alpha t^2$, i.e.
\be
\label{eq:wave_expa}
\wavelet(t) = \e^{\beta t + \gamma }
\sum_{n=0}^\infty \frac{\left( \alpha t^2  \right)^n}{n!}\,.
\ee
Each term in Eq.~\eqref{eq:wave_expa} 
can be analytically integrated,
leading to a well-defined solution 
of the Fourier counterpart.
In particular,
\be
\label{eq:fdw_expa_compute}
\wavelet(f) = \e^\gamma\,
\sum_{n=0}^\infty 
\int_0^{\tau} \frac{\left( \alpha t^2  \right)^n}{n!}\,
\e^{(\beta -2\pi\i f)t}\,\d t\,,
\ee
from which it follows
\be
\label{eq:fdw_expa}
\wavelet(f) = \frac{\e^{\gamma }}{\sqrt{\pi}} \sum_{n=0}^\infty 
\left(4\alpha\right)^n\, \Gamma{\left(n+\half\right)}\,
\frac{G_{2n} (-\zeta \tau )-1}{\zeta ^{2n+1}}\,,
\ee
where 
$\Gamma(n)$ is the gamma function
and $G_n (x)$ corresponds to
\be
\label{eq:gamma_approx}
G_n(x) = \e^{-x}\,\sum_{k=0}^{n}\frac{x^k}{k!}\,.
\ee
We observe that $G_n\to 1$ for $n \to \infty$.
Limiting the series to $n=0$,
Eq.~\eqref{eq:fdw_expa} leads to
the damped sinusoidal case, i.e. Eq.~\eqref{eq:wave_damped}.
In our implementation,
we use Eq.~\eqref{eq:fdw_expa}
as approximation of $\wavelet(f)$
for $|\alpha|\tau^2\lesssim 0.1$
accounting up to $n=4$.

\section{FM approximation}
\label{app:fmapprox}

In this appendix, we discuss the approximation
performed in order to reach an analytical form for the 
FM effects in terms of $\wavelet(f)$, i.e. Eq.~\eqref{eq:fdwavelet}.

Let us start considering a generic non-modulated wavelet $\wavelet(t)$,
as the one in Eq.~\eqref{eq:tdwavelet}. 
This term can be decomposed in amplitude and phase,
analogously to Eq.~\eqref{eq:hlm}, from which we
can compute the frequency, that reads
\be
\label{eq:fh}
\omega_\wavelet(t) = - 2 \ima t - \imb \,.
\ee
In order to include damped FMs,
we generalize the notion of $\wavelet$
introducing $\wavefm$, such that
\be
\label{eq:fhfm}
\omega_\wavefm(t) = \omega_\wavelet(t) - \Delta_{\rm fm} \e^{-\Gamma_{\rm fm} t}\sin(\Omega_{\rm fm} t + \phi_{\rm fm})\,,
\ee
where $\Delta_{\rm fm},
\Gamma_{\rm fm},
\Omega_{\rm fm},
\phi_{\rm fm}
\in\mathbb{R}$ 
define the modulation, i.e.
the frequency displacement $\Delta_{\rm fm}\ge 0$, 
the inverse damping time $\Gamma_{\rm fm}$,
the modulation frequency $\Omega_{\rm fm}$ 
and the initial phase $\phi_{\rm fm}$.
Integrating Eq.~\eqref{eq:fhfm},
the frequency-modulated wavelet $\wavefm(t)$
can be rewritten in the time-domain as
\be
\label{eq:fmwavelet2}
\wavefm(t)=
\wavelet(t;\alpha,\beta,\gamma,\tau)\,\e^{-\i
	\fm(t;\Delta_{\rm fm},\Gamma_{\rm fm},\Omega_{\rm fm},\phi_{\rm fm})}\,,
\ee
where $\fm(t)$ corresponds to 
\begin{widetext}
\be
\label{eq:fm}
	\fm(t) 
=\frac{ \Delta_{\rm fm}\,\e^{-\Gamma_{\rm fm} t}}{\Gamma_{\rm fm}^2+\Omega_{\rm fm}^2}\,\big[
\Gamma_{\rm fm} \sin(\Omega_{\rm fm} t + \phi_{\rm fm})
+\Omega_{\rm fm} \cos(\Omega_{\rm fm} t + \phi_{\rm fm})\big] - \fmzero\,,
\ee
\end{widetext}
with 
\be
\label{eq:fm_extraphase}
\fmzero =\frac{\Delta_{\rm fm}}{\Gamma_{\rm fm}^2 + \Omega_{\rm fm}^2}(\Gamma_{\rm fm}\sin\phi_{\rm fm} + \Omega_{\rm fm}\cos\phi_{\rm fm})\,.
\ee
Notice that $\fm(t)\in \mathbb{R}$ and $\e^{-\i\fm(t)}$
is a unitary complex factor for every given $t$. 

Due to the oscillatory nature of $\fm(t)$,
the frequency-domain wavelet
$\wavefm(f)$ cannot be 
analytically computed using Gaussian integration rules.
Then, 
we rewrite $\fm(t)$ in terms of exponential functions,
 \be
 \label{eq:fmfexp}
 \fm(t) = \frac{\i\Delta_{\rm fm}}{2|\beta_{\rm fm}|^2}\left(\beta_{\rm fm}^* \e^{-\beta_{\rm fm} t -\i\phi_{\rm fm}}-\beta_{\rm fm} \e^{-\beta_{\rm fm}^* t+\i\phi_{\rm fm}}\right) - \fmzero\,,
  \ee
  where $\beta_{\rm fm} = \Gamma_{\rm fm} +\i\Omega_{\rm fm}$.
Subsequently, we expand the exponential 
$\e^{-\i\fm}$, i.e.
 \be
\label{eq:fmapprox2}
\e^{-\i\fm(t)}=  \sum_{n=0}^\infty \frac{\left[-\i\fm(t)\right]^n}{n!} \,.
\ee
Combining Eq.~\eqref{eq:fmwavelet2},
~\eqref{eq:fmfexp} and
\eqref{eq:fmapprox2},
we can write $\wavefm(t)$ in terms of $\wavelet(t)$
and perform an analytical Fourier transform.
In particular,
\be
\label{eq:fmwave_fd}
\wavefm(f) = \e^{\i \fmzero}\,\sum_{n=0}^\infty \, \left(\frac{\Delta_{\rm fm}}{2|\beta_{\rm fm}|^2}\right)^n\,
\frac{w_{n}(f)}{n!}\,,
\ee
where
\be
\label{eq:fmwave_hfd}
w_{n}(f) = \sum_{k=0}^n {n\choose k} \,(\beta_{\rm fm}^*)^k\,(-\beta_{\rm fm})^{n-k}\,\wavelet(f;\alpha,\beta_{n,k},\gamma_{n,k},\tau)\,,
\ee
with 
\be
\label{eq:fmwave_hfd_coeffs}
\begin{split}
	\beta_{n,k} &=\beta -k\beta_{\rm fm} - (n-k)\beta_{\rm fm}^* \\
	&=\beta-n\Gamma_{\rm fm} + \i(n-2k)\Omega_{\rm fm}\,, \\
	\gamma_{n,k} &= \gamma_{\rm fm}+\i (n-2k)\phi_{\rm fm}\,.
\end{split}
\ee
and $\{\alpha,\beta,\gamma,\tau\}$ are the parameters of the 
corresponding non-modulated wavelet.

Eq.~\eqref{eq:fmwave_fd} generates several 
Fourier contributions
centered around the frequencies 
$\imb \pm n \Omega $,
as expected from FM effects.
A second order approximation gives good agreement for small
modulation indices, i.e. $\Delta/ \Omega \ll 1$; 
however, when $\Delta$ is comparable to $\Omega$,
additional terms need to be taken into account for an accurate 
description.
The maximum order of approximation $n_{\rm max}$ 
is estimated using an empirical rule of thumb,
$n_{\rm max} \approx 2\,\left( 1+ {\Delta}/{\Omega} \right)$.

\section{Choices for wavelet composition}
\label{app:wavecomponent}

The first contribution, $\wavelet_{\rm fus}$,
corresponds to the fusion of the NS cores. This term 
is modeled with a Gaussian wavelet (i.e. $\i\beta\in\mathbb{R}$), 
with initial amplitude, frequency and phase defined by the values at merger,
respectively $A_{\rm mrg}$, $f_{\rm mrg}$ and $\phi_{\rm mrg}$.
The width of the amplitude is fixed as follows, 
\be
\label{eq:spiral_rea}
\realabel{\rm fus} = \frac{\log(A_{0}/A_{\rm mrg})}{t^2_{0}}\,,
\ee
while the frequency slope $\imalabel{\rm fus}$ is 
directly estimated from NR data.
The fusion wavelet $\wavelet_{\rm fus}$ is truncated at $t_{0}$.
It follows that
\be
\label{eq:spiralwavelet}
\begin{split}
	\wavelet_{\rm fus} (f) = \wavelet(f; \,\,	&\alpha=\realabel{\rm fus} - \i\imalabel{\rm fus}, \\
	&\beta = -2\pi \i f_{\rm mrg},\\
	& \gamma=\log(A_{\rm mrg})-\i\phi_{\rm mrg},\\
	&\tau = t_0, \\
	&\twshift=0)\,.
\end{split}
\ee 

Subsequently, we include an intermediate wavelet $\wavefm_{\rm bnc}$
that characterizes the bounce of the remnant after
the collision of the NS cores,
corresponding to the time interval $[t_{0}, t_{1}]$.
The initial amplitude is determined in order to match
the $A_{0}$ and the phase is computed from 
the wavelet $\wavelet_{\rm fus}$ including an additional phase-shift
$\phi_{\rm PM}$, shown by NR simulations~\cite{Kastaun:2016yaf,Dietrich:2018phi}, i.e.
\be
\label{eq:phi0_bounce}
\phi_{\rm bnc}= \phi_{\rm mrg} + \phi_{\rm PM} +2\pi f_{\rm mrg} t_{0} + \imalabel{\rm fus} t_{0}^2\,.
\ee
The amplitude coefficients, $\realabel{\rm bnc}$ and 
$\reblabel{\rm bnc}$, are chosen such that the amplitude peaks 
in the first local amplitude maximum, i.e. $t_{1}$, with value $A_{1}$;
\begin{align}
\label{eq:reab_bounce}
\realabel{\rm bnc}& = \frac{\log(A_{0}/A_{1})}{(t_{1}-t_{0})^{2}}\,,\\
\reblabel{\rm bnc}& =\frac{2\log(A_{1}/A_{0})}{t_{1}-t_{0}}\,.
\end{align}
The frequency is kept constant with value $\imblabel{\rm bnc}=-2\pi \i f_{2}$.
Then, including FM effects as discussed in Sec.~\ref{sec:modulations},
we get
\be
\label{eq:recoilwavelet}
\begin{split}
	\wavefm_{\rm bnc} (f) = \wavelet(f; \,\,	&\alpha=\realabel{\rm bnc}, \\
	&\beta = \reblabel{\rm bnc} -2\pi \i f_{2},\\
	& \gamma=\log(A_{0})-\i\phi_{\rm bnc},\\
	&\tau = t_{1} - t_{0}, \\
	&\twshift= t_{0},\\
	& \Delta_{\rm fm} = \Delta_{\rm fm},\\
	& \Gamma_{\rm fm} = 0,\\
	& \Omega_{\rm fm} =  2\pi f_0,\\
		& \phi_{\rm fm} =  \phi_{\rm fm})\,.
\end{split}
\ee

After $t_1$, the remnant is strongly deformed and the 
quadrupolar radiation is affected by couplings with subdominant modes,
that introduce AMs. 
Physically, this phenomenon can be naively interpreted with
the presence of radial pulsation 
in the mass distribution of the remnant object~\cite{Bauswein:2015yca}.
We limit ourselves to the modeling of AMs in the region $[t_1, t_3]$
taking into account the coupling with the $(2,0)$ mode,
analogously to Ref.~\cite{Breschi:2019srl}.
This pulsating portion of signal can be approximated using a 
wavelet $\wavegen_{\rm pul}$ of the form,
\be
\label{eq:deformwavelet_td}
\wavegen_{\rm pul} (t) = A_1 \left[1 -\Delta_{\rm am} \sin^2\left(\pi f_0 t\right) \right] \e^{\left[\reblabel{\rm pul} -2\pi \i f_2\right] t -\i  \phi_{\rm pul}}\,,
\ee
where the initial amplitude and phase are chosen to match
values of $\wavefm_{\rm bnc}$ at $t_{1}$, in particular the phase
$\phi_{\rm pul}$ corresponds to
\be
\label{eq:phi0_deform}
\phi_{\rm pul}= \phi_{\rm bnc} + 2\pi f_{2} (t_1-t_{0})\,,
\ee
the coefficient $\reblabel{\rm pul}$ is defined by the amplitudes $A_{1,3}$ as 
\be
\label{eq:reb_def}
\reblabel{\rm pul}=\frac{\log\left({A_3}/{A_1}\right)}{t_3-t_1}\,,
\ee
and the coefficient $\Delta_{\rm am}$ defines the magnitude of AMs,
\be
\label{eq:mu_defs}
\Delta_{\rm am} = 1 - \frac{A_2}{A_1} \left(\frac{A_1}{A_3}\right)^{\frac{t_2-t_1}{t_3-t_1}}=1 - \frac{A_2}{\sqrt{A_1\,A_3}} \,,
\ee
where we made use of 
the definition of $t_i$ (Sec.~\ref{sec:nodes}) 
in the second equality.
Then, Eq.~\eqref{eq:deformwavelet_td} can be rewritten in terms of 
frequency-domain wavelets, Eq.~\eqref{eq:fdwavelet}, as
\be
\label{eq:deformwavelet}
\begin{split}
	\wavegen_{\rm pul} (f) =\left(1-\frac{\Delta_{\rm am}}{2}\right) \wavelet(f; \,\,	&\alpha=0, \\
	&\beta = \reblabel{\rm pul} -2\pi \i f_{2},\\
	& \gamma=\log(A_{1})-\i\phi_{\rm pul},\\
	&\tau = t_{3} - t_{1}, \\
	&\twshift= t_{1},\\
	& \Delta_{\rm fm} = \Delta_{\rm fm},\\
	& \Gamma_{\rm fm} = \Gamma_{\rm fm} ,\\
	& \Omega_{\rm fm} =  2\pi f_0,\\
	& \phi_{\rm fm} =  \phi_{\rm fm})\\
	+\,\,\frac{\Delta_{\rm am}}{4}\,\,  \wavelet(f; \,\,	&\alpha=0, \\
	&\beta = \reblabel{\rm pul} -2\pi \i (f_{2}-f_0),\\
	& \gamma=\log(A_{1})-\i\phi_{\rm pul},\\
	&\tau = t_{3} - t_{1}, \\
	&\twshift= t_{1},\\
	& \Delta_{\rm fm} = \Delta_{\rm fm},\\
	& \Gamma_{\rm fm} = \Gamma_{\rm fm} ,\\
	& \Omega_{\rm fm} =  2\pi f_0,\\
	& \phi_{\rm fm} =  \phi_{\rm fm})\\
	+\,\,\frac{\Delta_{\rm am}}{4} \,\, \wavelet(f; \,\,	&\alpha=0, \\
	&\beta = \reblabel{\rm pul} -2\pi \i (f_{2}+f_0),\\
	& \gamma=\log(A_{1})-\i\phi_{\rm pul},\\
	&\tau = t_{3} - t_{1}, \\
	&\twshift= t_{1},\\
	& \Delta_{\rm fm} = \Delta_{\rm fm},\\
	& \Gamma_{\rm fm} = \Gamma_{\rm fm} ,\\
	& \Omega_{\rm fm} =  2\pi f_0,\\
	& \phi_{\rm fm} =  \phi_{\rm fm})\,.
\end{split}
\ee

Subsequently, we model the signal with a damped
tail related to the quadrupolar deformations of the rotating remnant.
The corresponding wavelet $\wavefm_{\rm peak}$
is modeled in the range $[t_{3}, t_{\rm coll}]$.
If the remnant is a stable NS configuration, $ t_{\rm coll}\to\infty$.
The initial amplitude and phase are chosen to match the 
values of $\wavegen_{\rm pul}$ at $t_{3}$,
\begin{align}
\label{eq:phi_tail}
\phi_{\rm peak}& = \phi_{\rm pul} + 2\pi f_2 (t_{3}-t_{1}) \,,\\
A_{\rm peak}& =A_{3}\,.
\end{align}
The frequency evolution is characterized by the typical
$f_2$ peak, i.e. $\imblabel{\rm peak}=-2\pi f_2$,
with a non-vanishing slope $\imalabel{\rm peak}$
(also referred as $\adrift$ in the manuscript 
to lighten the notation).
Then,
\be
\label{eq:tailwavelet}
\begin{split}
	\wavefm_{\rm peak} (f) = \wavelet(f; \,\,	&\alpha= - \i\Im{(\alpha_{\rm peak})}, \\
	&\beta =\reblabel{\rm peak} -2\pi \i f_{2},\\
	& \gamma=\log(A_{3})-\i\phi_{\rm peak},\\
	&\tau = t_{\rm coll} - t_{3}, \\
	&\twshift= t_{3},\\
	& \Delta_{\rm fm} = \Delta'_{\rm fm},\\
	& \Gamma_{\rm fm} = \Gamma_{\rm fm} ,\\
	& \Omega_{\rm fm} =  2\pi f_0,\\
	& \phi_{\rm fm} =  \phi_{\rm fm})\,,
\end{split}
\ee 
where $\Delta'_{\rm fm} = 
\Delta_{\rm fm}\,\exp[\Gamma_{\rm fm}(t_3-t_1)]$.

When $t_{\rm coll}$ is finite and the remnant collapse into BH, 
NR simulations show an increasing frequency and a damping amplitude
similarly to a BH ringdown.
This evolution can be captured with the inclusion of an additional wavelet component, 
i.e. $\wavelet_{\rm coll}$.
However, this contribution is expected to be relatively weak in terms of 
GW luminosity with respect to the previous dynamics~\cite{Zappa:2017xba}.
Moreover, the characteristic BH frequencies 
for this kind of systems lie in a very high frequency range,
roughly $\gtrsim 6$~kHz~\cite{Berti:2014fga}, 
where the sensitivities of 
the detectors are generally poor.
It follows that the collapse portion of the signal is
expected to have negligible effect on the overall GW power and,
for these reasons, we approximate $\wavelet_{\rm coll}= 0$.

The overall model includes 17 parameters:
the merger amplitude $A_{\rm mrg}$ and frequency $f_{\rm mrg}$;
the frequency drift at merger $\imalabel{\rm fus}$;
the characteristic PM frequencies $f_2$ and $f_0$;
the frequency drift $\imalabel{\rm peak}$
(or $\adrift$);
the time of the first nodal point $t_0$ 
and the corresponding phase-shift $\phi_{\rm PM}$;
the amplitude values at the different nodal points $\{A_i\}$,
for $i=0,1,2,3$; 
the inverse damping time of the Lorentzian tail $\imblabel{\rm peak}$;
the time of collapse $t_{\rm coll}$;
and the FM properties, i.e. $\Delta_{\rm fm}$ and $\Gamma_{\rm fm}$ and$\phi_{\rm fm}$.
Finally, we observe that $\wavefm_{\rm bnc} (f)$, $\wavegen_{\rm pul} (f) $
and $\wavefm_{\rm peak} (f)$ are chosen to be identically
zero for $\Lambda_1 = \Lambda_2 =  0$.

\section{Viscosity impact on frequency drift}
\label{app:viscosity}

\begin{figure}[t]
	\centering 
	\includegraphics[width=0.49\textwidth]{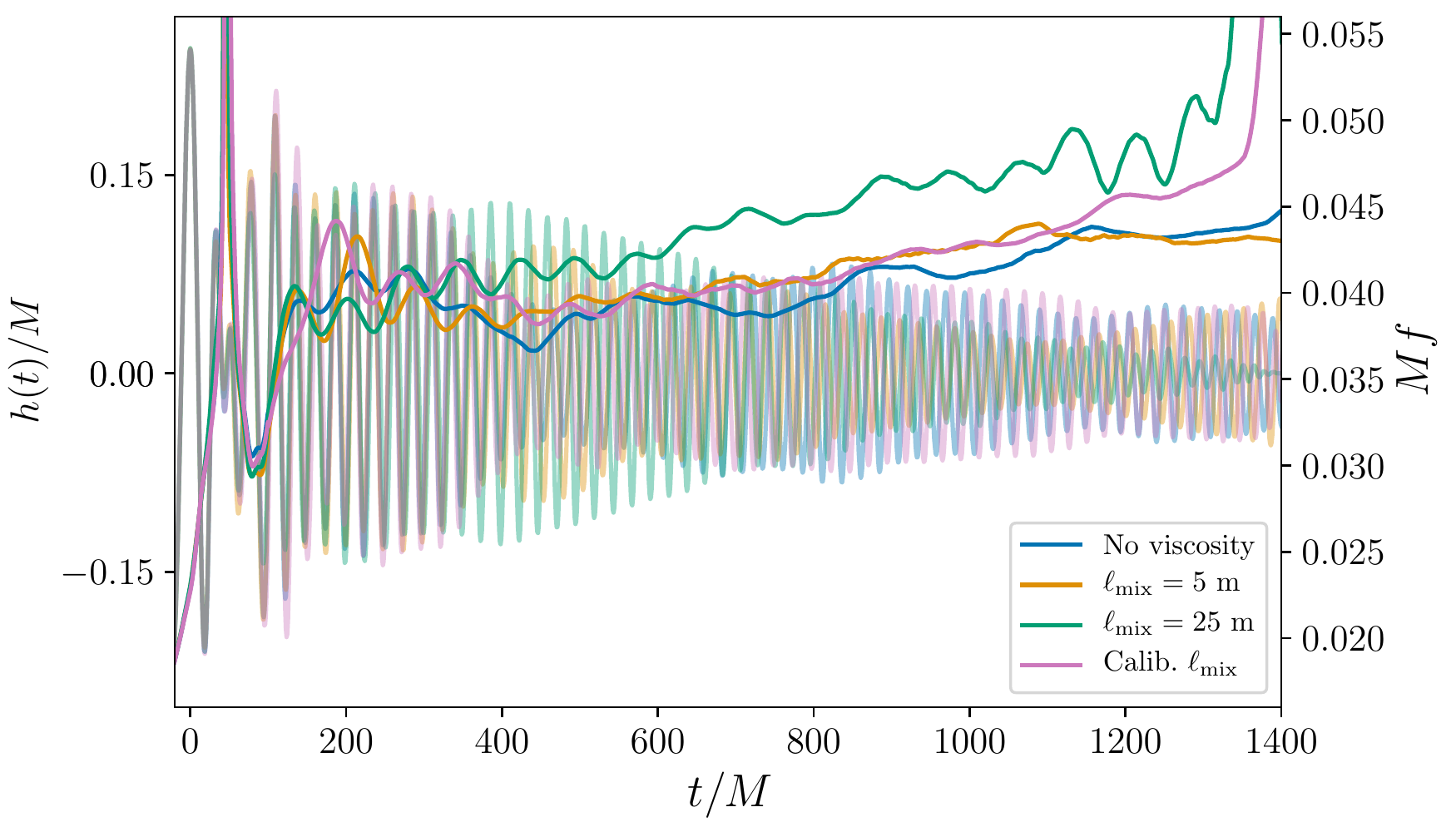}
	\caption{GW data extracted from NR simulations
					with neutrino reabsorption 
					of a BNS merger with $M=2.7~\Msun$, $q=1$ and LS220 EOS~\cite{Radice:2017zta,Radice:2018pdn}.
					Blue curve refers to data with no turbolent viscosity,
					orange curve refer to fixed $\ell_{\rm mix}=5~{\rm m}$,
					green curve refers to fixed $\ell_{\rm mix}=25~{\rm m}$ 
					and
					purple curve refers to $\ell_{\rm mix}$ 
					calibrated on \cite{Kiuchi:2017zzg}.
					Solid lines show the GW frequency $Mf$
					and shaded lines show the GW waveform $h/M$.
					The instant $t=0$ corresponds to the merger.}
	\label{fig:visc}
\end{figure}

In this Appendix,
we show the effects of different viscosity schemes 
on the dynamical evolution of the GW frequency.
This discussion aims to motivate the introduction
of the frequency drift $\adrift$ as free parameter.

Figure~\ref{fig:visc} shows three NR simulations
extracted from \cite{Radice:2017zta,Radice:2018pdn}
and computed with identical grid resolution.
The data correspond to a BNS system with $M=2.7~\Msun$ and $q=1$ with matter properties described by the same EOS, i.e. LS220~\cite{Lattimer:1991nc}.
Moreover, all cases include neutrino reabsorption scheme~\cite{Radice:2016dwd}.
The blue curves refer to binaries with
no turbolent viscosity.
The orange and green curves include turbolent viscosity
with a fixed mixing length
respectively equal to $\ell_{\rm mix}=5~{\rm m}$ 
and $\ell_{\rm mix}=25~{\rm m}$.
The mixing length $\ell_{\rm mix}$ represents the
characteristic scale over which
turbulence acts~\cite{Radice:2017zta}.
Finally,  the purple data are simulated with a
turbolent viscosity scheme calibrated on 
high-resolution magneto-hydro-dynamical simulations
of BNS mergers~\cite{Kiuchi:2017zzg}.

Over the domain $t/M\lesssim 300 $, 
the different cases show a similar behavior.
However, for later times,
the frequency drift significantly differs,
with a more pronounced slope for the 
$\ell_{\rm mix}=25~{\rm m}$ case.
The same binary is the one that shows the earliest BH collapse.
Notably, as shown in \cite{Radice:2017zta},
the frequency slope is softer for $\ell_{\rm mix}=50~{\rm m}$
with respect to the $\ell_{\rm mix}=25~{\rm m}$ case;
however, the assumption $\ell_{\rm mix}=50~{\rm m}$
appears to be physically disfavored
from studies of magnetorotational instability turbolence
in BNS simulations~\cite{Kiuchi:2017zzg}.
On the other hand, the simulation
with calibrated $\ell_{\rm mix}$ shows
an initial trend similar to the $\ell_{\rm mix}=5~{\rm m}$ case;
later, for $t/m\gtrsim1000$,
the two GW frequencies differ
and the calibrated-viscosity case shows a BH collapse.
Interestingly, the binaries with the 
steepest 
frequency drifts
tend to generate shorter GW bursts due to earlier BH collapse.
These physical effects
cannot be described by the binary properties only
(i.e. masses, spins and tides).
Thus, it is necessary to rely on additional 
coefficients that aim to characterize the
physical information on the matter 
dynamics encoded in the PM transients.

\section{Additional parameters for {\oldmodel}}
\label{app:nrpm_newpars}

In \cite{Breschi:2021xrx} and in Sec.~\ref{sec:fbar},
we introduced the additional parameters $\{\alpha,\beta,\phi_{\rm PM}\}$
for the {\oldmodel} model~\cite{Breschi:2019srl}.
This Appendix aims to expand this discussion,
specifying the role of each term
with reference to \cite{Breschi:2019srl}.

The additional phase $\phi_{\rm PM}$
affects the phase evolution of {\oldmodel} 
introducing a phase discontinuity in $t_0$,
as previously discussed for {\model}.
The damping time $\alpha$, defined in Eq.~(11) of \cite{Breschi:2019srl},
is promoted to additional parameter,
since it improves the fitting of the characteristic peak for {\oldmodel}.
This is also motivated by the relation between the time of collapse 
and the high-density EOS properties~\cite{Radice:2016rys,
	Prakash:2021wpz,
Fujimoto:2022xhv}.
Finally, the $\beta$ parameter is inspired by the 
template model introduced in \cite{Easter:2020ifj}
and it takes into account linear deviations from the 
peak frequency $f_2$. In particular,
it modifies Eq.~(7d) of \cite{Breschi:2019srl}
as
\be
\label{eq:app:beta}
\hat \omega(\hat t>\hat t_3) = \hat \omega_2\cdot 
\left[ 1 + \beta (\hat t-\hat t_3)\right]\,.
\ee
As discussed for {\model} and in App.~\ref{app:viscosity},
general PM GW transients show non-vanishing frequency slope 
and this feature appears to correlate with the viscosity scheme employed
in the NR simulation.

\section{Quasiuniversal relations of type $f_2(R)$}
\label{app:eos_qur}

\begin{figure*}[t]
	\centering 
	\includegraphics[width=0.49\textwidth]{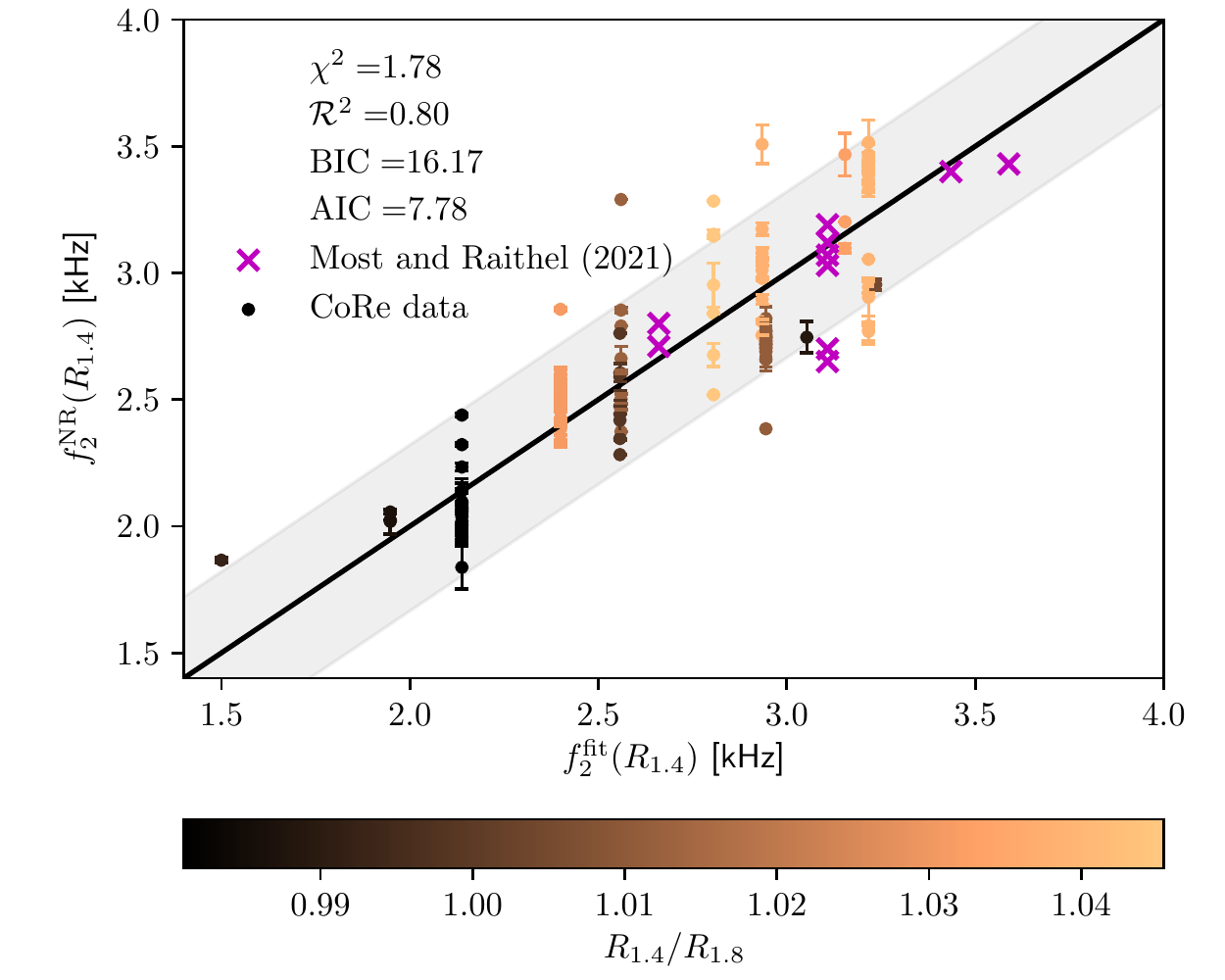}
	\includegraphics[width=0.49\textwidth]{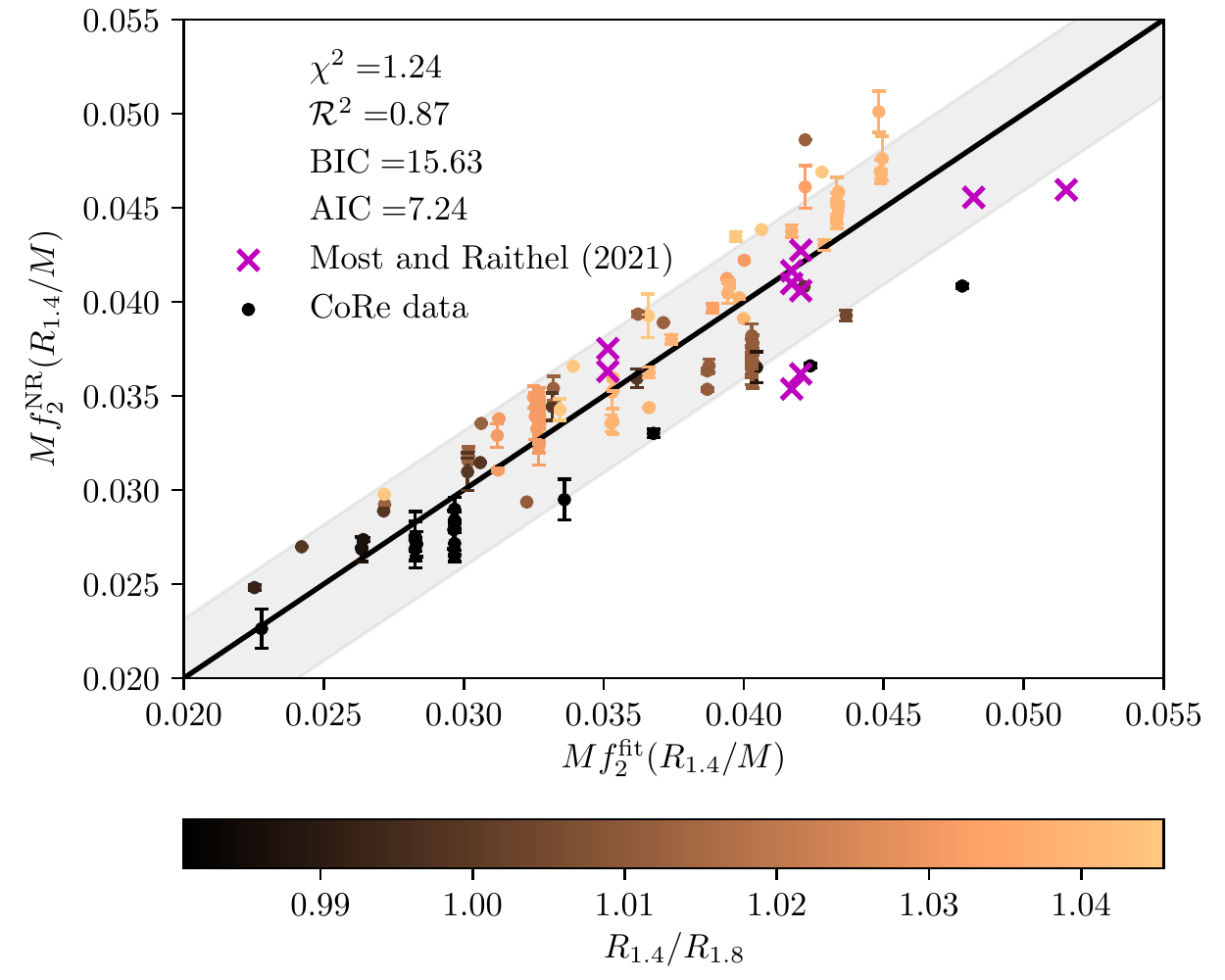}\\
	\includegraphics[width=0.49\textwidth]{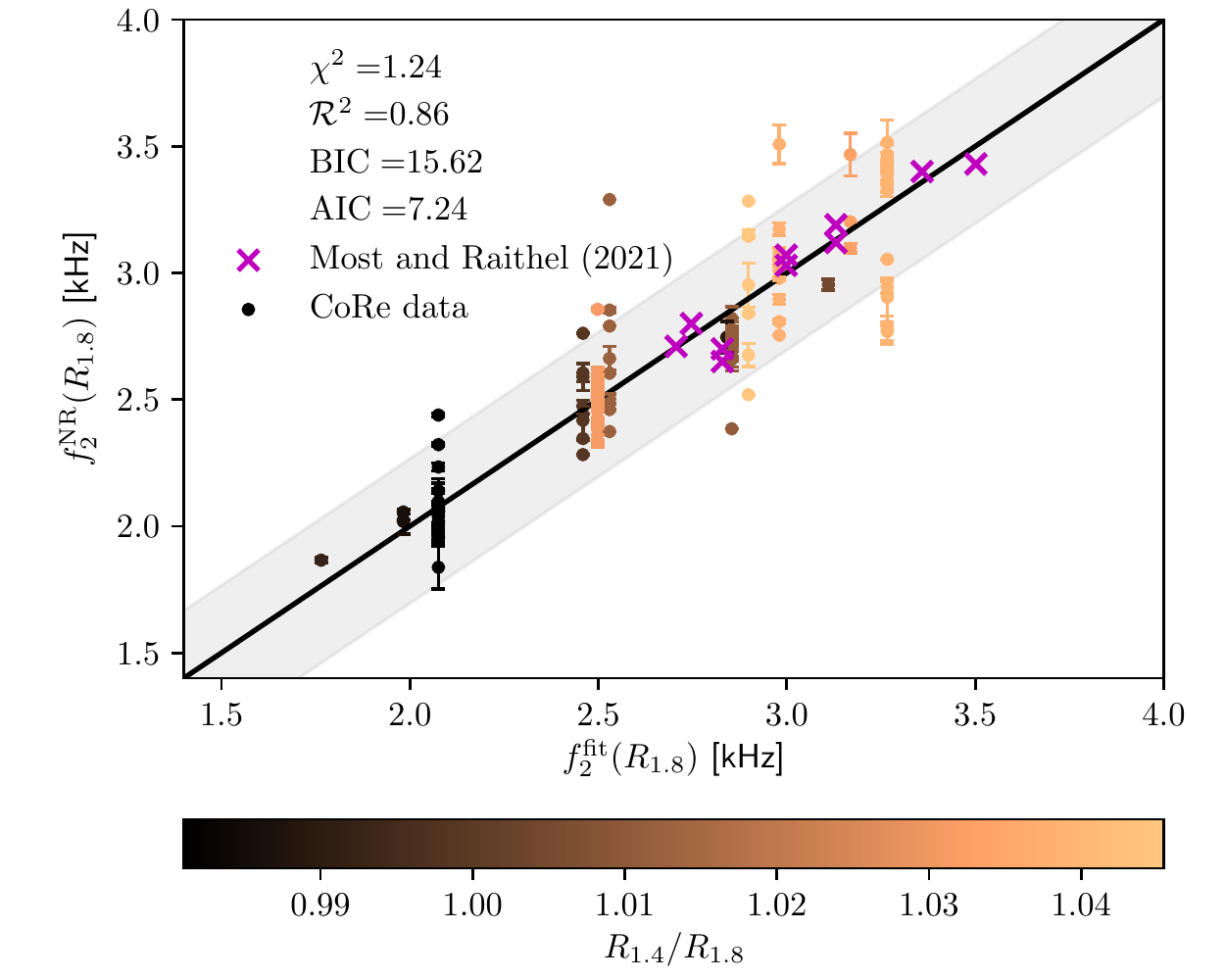}
	\includegraphics[width=0.49\textwidth]{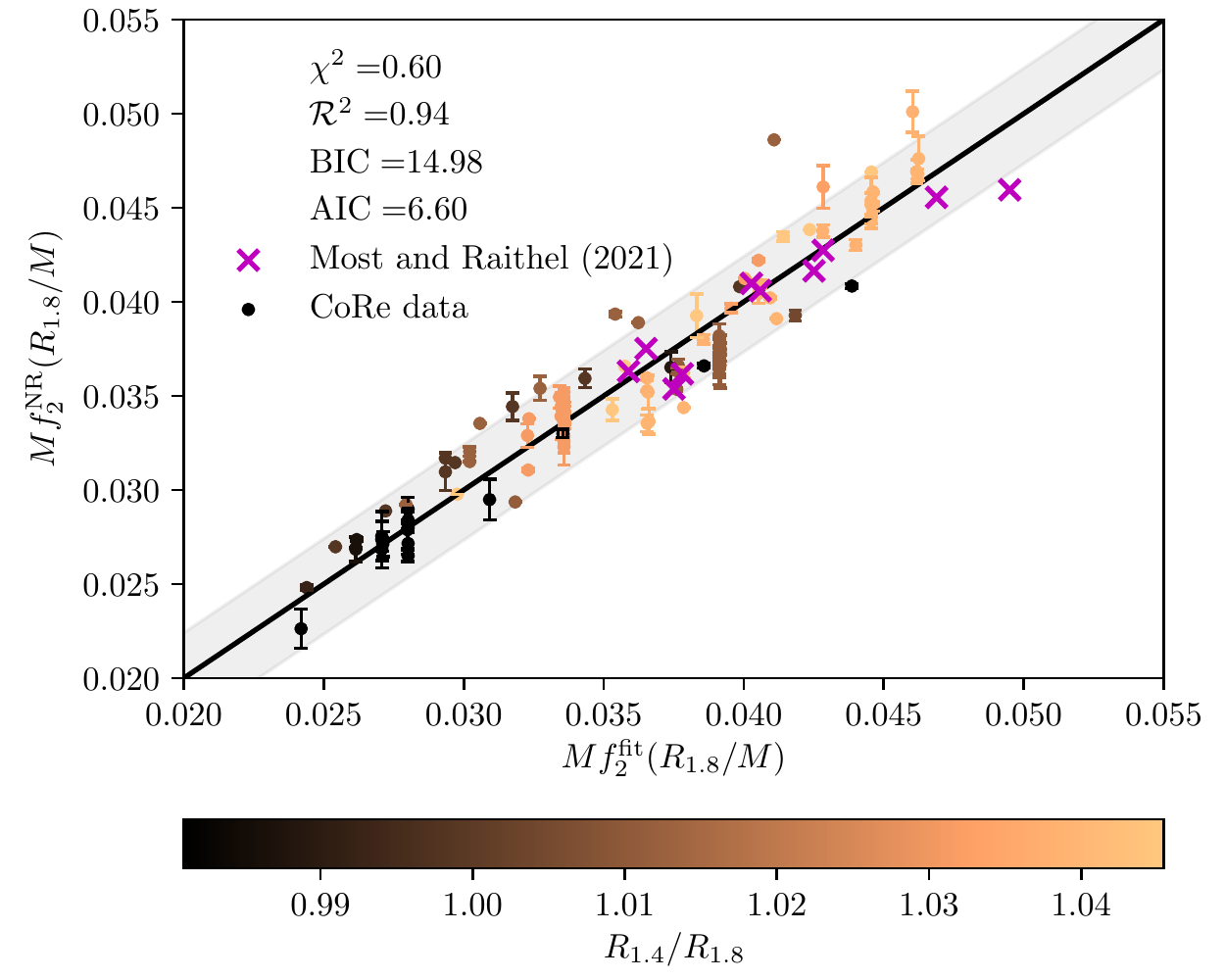}\\
	\caption{Predicted values from the calibrated relations 		
					Eq.~\eqref{eq:app:fit1}
					compared to the
					respective NR observed quantities. 
					Top panels show $X=1.4$ and bottom panels show $X=1.8$.
					Left panels show non-mass-scaled $f_2$
					and right panels show mass-scaled dimensionless $Mf_2$.
					The diagonal (black line) represents the case in which
					predictions and observations match and the gray area is
					the 90\% credibility level.
					The {\core} data are reported with circles colored according to
					$R_{1.4}/R_{1.8}$
					and magenta crosses are the data extracted from \cite{Most:2021ktk}.
        }
	\label{fig:f2_of_R}
\end{figure*}

\begin{figure*}[t]
	\centering 
	\includegraphics[width=0.49\textwidth]{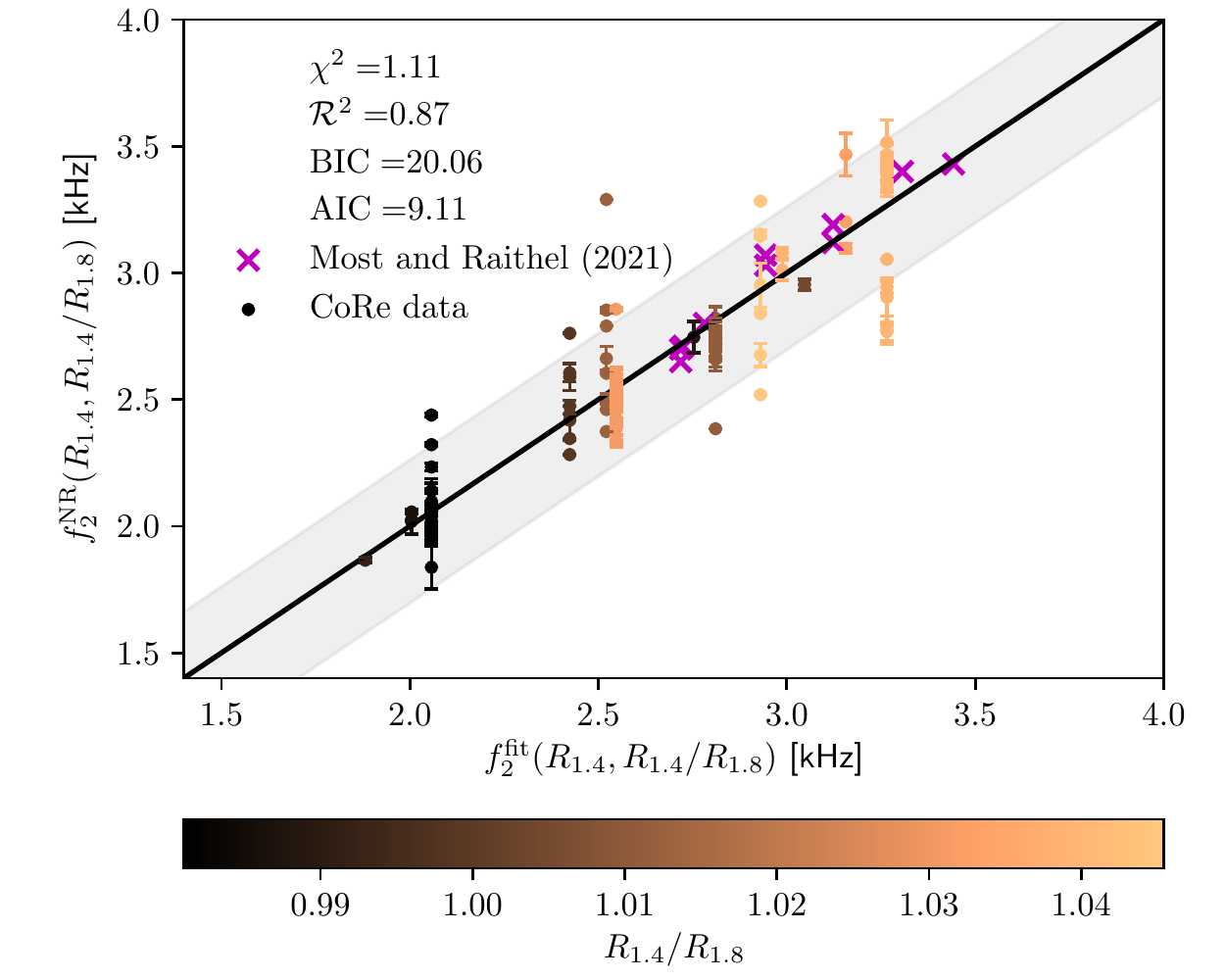}
	\includegraphics[width=0.49\textwidth]{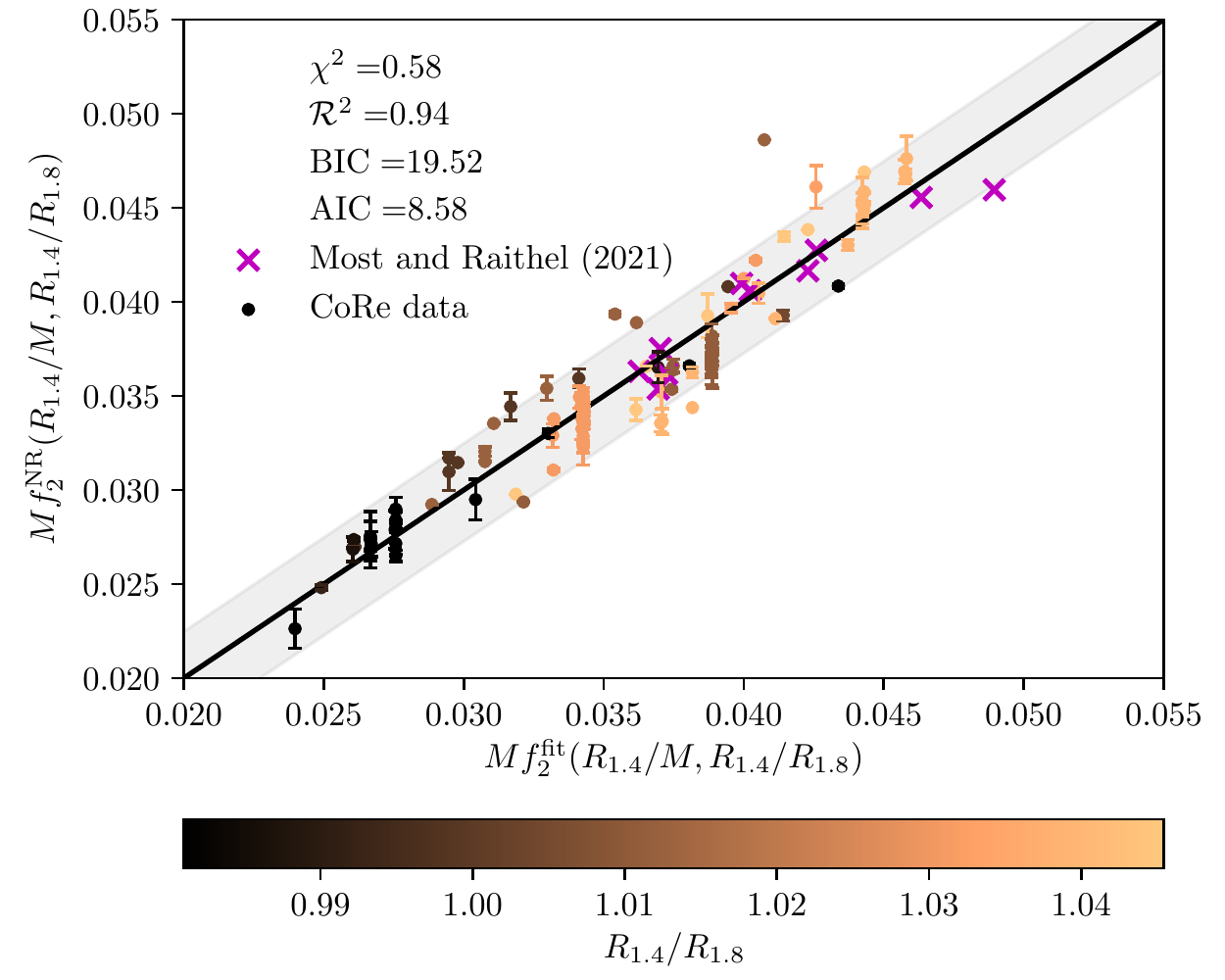}\\
	\includegraphics[width=0.49\textwidth]{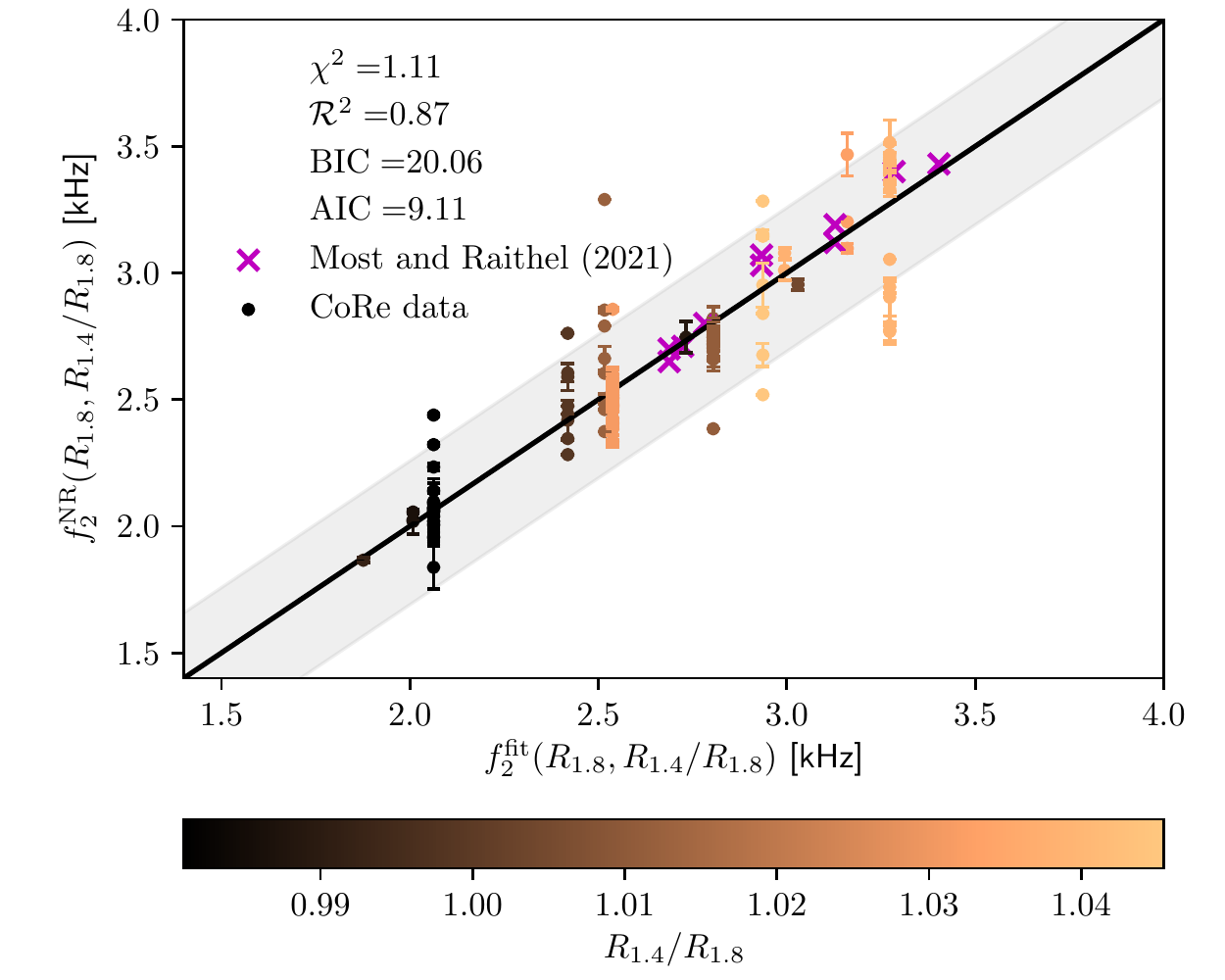}
	\includegraphics[width=0.49\textwidth]{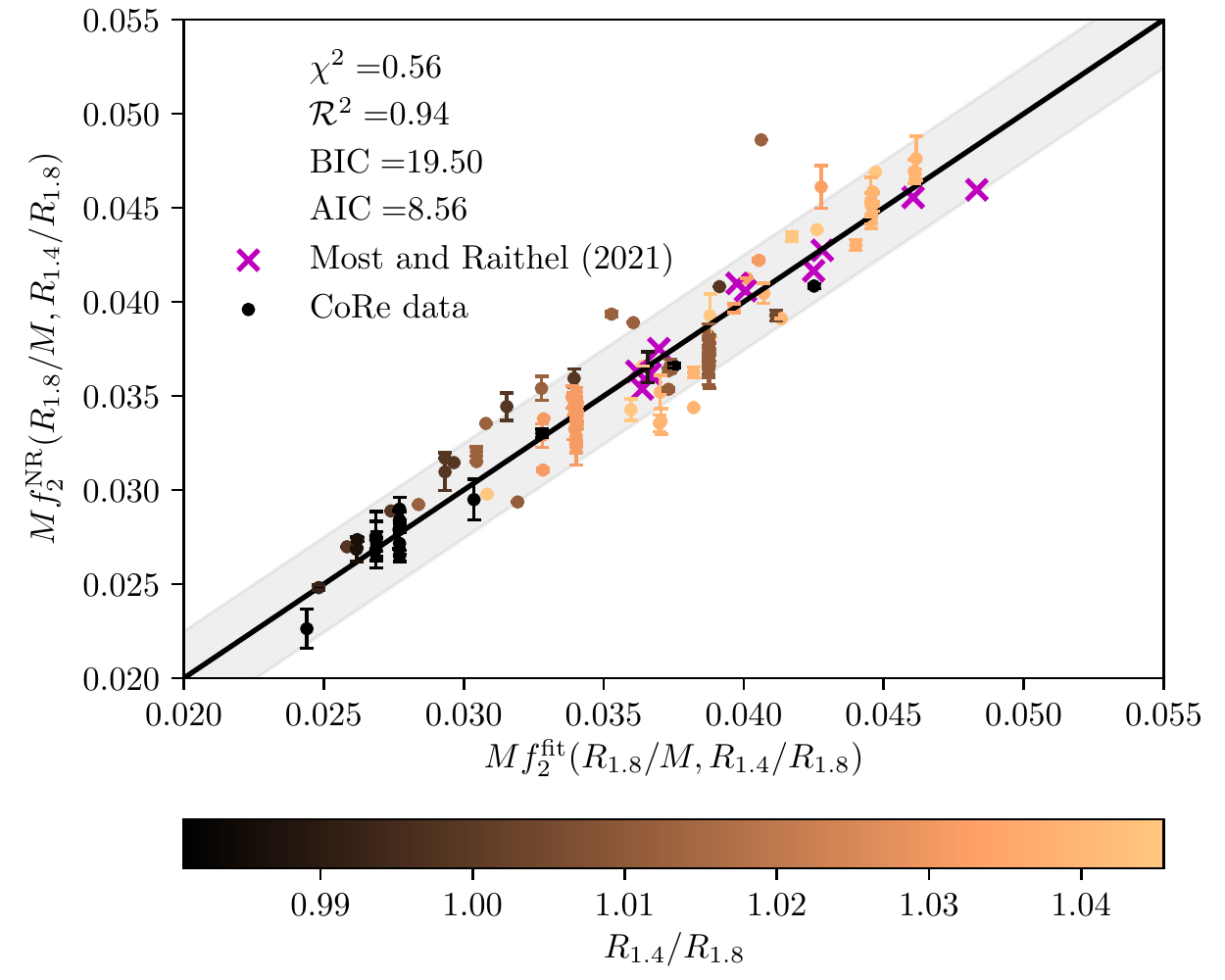}\\
	\caption{Predicted values from the calibrated relations 		
		Eq.~\eqref{eq:app:fit2} compared to the
		respective NR observed quantities. 
		Top panels show $X=1.4$ and bottom panels show $X=1.8$.
		Left panels show non-mass-scaled $f_2$
		and right panels show mass-scaled dimensionless $Mf_2$.
		The diagonal (black line) represents the case in which
		predictions and observations match and the gray area is
		the 90\% credibility level.
		The {\core} data are reported with circles colored according to
		$R_{1.4}/R_{1.8}$
		and magenta crosses are the data extracted from \cite{Most:2021ktk}.}
	\label{fig:f2_of_R_corr}
\end{figure*}

\begin{table}[t]
	\centering    
	\caption{Summary of the calibrated relations for the PM peak 
					frequency $f_2$ as function of the NS radii $R_{1.4}$ and
					$R_{1.8}$.
					The first column shows the calibrated quantity of interest;
					the calibrated values of the empirical coefficients are 
					reported from the second to the fifth column.}
	\resizebox{0.49\textwidth}{!}{
		\begin{tabular}{c|cccc}        
			\hline
			\hline
			$Q^{\rm fit}$ & $a_0$ & $a_1$ & $a_2$ & $a_3$  \\
			\hline
			\hline
			$f_2(R_{1.4})$ & $5.42$ & $0.0449$ & $-0.0198$ & -- \\
			$f_2(R_{1.8})$ & $11.5$ & $-0.990$ & $0.0233$ & -- \\
			\hline 
			$Mf_2\left({R_{1.4}}/{M}\right)$ & $0.2$ & $-0.0762$ & $ 0.0078$ & -- \\ 
			$Mf_2\left({R_{1.8}}/{M}\right)$ & $0.236$ & $ -0.103$ & $0.0125$ & -- \\
			\hline
			\hline
			$f_2\left(R_{1.4}, {R_{1.4}}/{R_{1.8}}\right)$ & $6.4$ & $-1.33$ & $0.0381$ & $6.97$ \\
			$f_2\left(R_{1.8}, {R_{1.4}}/{R_{1.8}}\right)$ & $9.99$ & $-1.24$ & $0.0349$ & $2.76$\\
			\hline 
			$Mf_2\left({R_{1.4}}/{M}, {R_{1.4}}/{R_{1.8}}\right)$ & $0.162$ & $-0.115$ & $0.0145$ & $0.0919$ \\
			$Mf_2\left({R_{1.8}}/{M}, {R_{1.4}}/{R_{1.8}}\right)$ & $0.213$ & $-0.105$ & $ 0.013$ & $0.0241$ \\
			\hline
			\hline
	\end{tabular}}
	\label{tab:f2R_fits}
\end{table}

In this appendix, we discuss the quasiuniversal relations 
between the PM peak frequency $f_2$ and the NS radius 
at fiducial values in light of the results of 
\cite{Most:2021ktk,Raithel:2022orm}.
In particular, 
we calibrate the relations $f_2(R_{1.4})$ and 
$f_2(R_{1.8})$ including the {\core} data~\cite{Dietrich:2018phi,Radice:2018pdn,
	Bernuzzi:2020txg,Nedora:2020pak,Prakash:2021wpz}, 
where $R_{1.4}$ ($R_{1.8}$) 
is the radius of a $1.4~\Msun$ ($1.8~\Msun$) NS
computed from the TOV equations.

Following \cite{Raithel:2022orm},
we employ a quadratic relation
for the calibration of the PM peak as function
of the NS radius, including linear corrections in the 
ratio ${R_{1.4}}/{R_{1.8}}$, i.e.
\begin{align}
\label{eq:app:fit1}
f_2(R_{X}) &= a_0 + a_1 R_{X} + a_2 R_{X}^2\,,\\
\label{eq:app:fit2}
f_2\left(R_{X}, \frac{R_{1.4}}{R_{1.8}}\right) &= a_0 + a_1 R_{X} + a_2 R_{X}^2+a_3\frac{R_{1.4}}{R_{1.8}}\,,
\end{align}
for $X= 1.4 , 1.8$,
where $f_2$ is measured in kHz
and $R_{X}$ in km.
Subsequently, 
we fit the NR data scaling the calibrated quantities by
the total mass $M$ of the system, i.e. 
$f_2\mapsto Mf_2$ and $R_{X}\mapsto R_{X}/M$.
Our final calibration set is composed by $65\%$ by binaries
with $R_{1.4}/R_{1.8}>1$. 
Table~\ref{tab:f2R_fits} shows the values of the calibrated coefficients $\{a_i\}$ for the different quantities.
Figure~\ref{fig:f2_of_R} and Figure~\ref{fig:f2_of_R_corr} show the NR data $f_2^{\rm NR}$ 
plotted against the predictions $f_2^{\rm fit}$ of the 
calibrated relation and the statistical quantities of 
interest; i.e.
the $\chi^2$ (defined in Sec.~\ref{sec:calib}),
the adjusted coefficient of determination $\mathcal{R}^2$,
the Bayesian information criterion (BIC)
and the Akaike information criterion (AIC).
For the computation of the BIC and the AIC,
we define a log-likelihood from Eq.~\eqref{eq:relres}
equal to $-\half\chi^2$. 

From our analysis, 
the calibrations performed with the mass-scaled quantities
show improved trends with respect to the analogous 
non-mass-scaled case.
This is due to the factorization of the total binary mass $M$, as expected by basic arguments in general relativity.
Moreover, the additional contribution $R_{1.4}/R_{1.8}$
appears to be more relevant for the calibration of low-density
properties, i.e. $f_2(R_{1.4})$, in agreement with \cite{Raithel:2022orm}.
However, the BIC and the AIC do not favor the introduction
of these additional term,
even if the $\chi^2$ of the calibrated relation $Mf_2(R_{1.4}/M)$
slightly decreases including $R_{1.4}/R_{1.8}$ in the fit.
We find that the most robust and reliable quasiuniversal relation
is the mapping $Mf_2(R_{1.8}/M)$,
which reinforces the hypothesis that PM quantities correlate 
with high-density EOS properties~\cite{Breschi:2021xrx}.
The differences between our results and the findings of \cite{Raithel:2022orm} 
might be related to the different size and composition of the
NR set, such as a different set of EOSs,
or to different definitions in the statistical quantities
\footnote{We verified that our results are stable when
					employing a Gaussian likelihood 
					or the standard Pearson's $\chi^2$ statistic.}.
Notably, simulations of the {\core} database show deviations comparable to 
the cases presented in \cite{Most:2021ktk,Raithel:2022orm}
that cannot be fully cured with the introduction
of the additional term $R_{1.4}/R_{1.8}$.

\end{document}